\documentclass[letterpaper,11pt]{article}
\pdfoutput=1
\usepackage{jheppub_mod}
\usepackage{graphicx, amsmath, amssymb, amstext,xcolor,txfonts}
\usepackage{relsize, float, multirow}
\usepackage{array, epstopdf, hyperref}
\usepackage{caption, subcaption, mathtools}
\usepackage{newtxmath,ulem}

\definecolor{seagreen}{rgb}{0.18, 0.55, 0.34}

\title{Utilizing localized fast radio bursts to constrain their progenitors and the expansion history of the Universe}

\author[a,1]{Sandeep Kumar Acharya,\note{Corresponding author.}}
\author[a,b,c]{Paz Beniamini}

\affiliation[a]{Astrophysics Research Center of the Open University, The Open University of Israel, Ra'anana, Israel}
\affiliation[b]{Department of Natural Sciences, The Open University of Israel, P.O Box 808, Ra’anana 4353701, Israel}
\affiliation[c]{Department of Physics, The George Washington University, 725 21st Street NW, Washington, DC 20052, USA}

\emailAdd{sandeepa@openu.ac.il}
\date{\today}

\abstract{Fast radio bursts (FRBs) are increasingly being used for cosmological applications such as measuring the Hubble constant and baryon abundance. The increasing number of localized FRBs and precise measurement of dispersion measure (DM) make them a suitable probe for such an approach. We use a sample of  110 localized FRBs as well as a small sub-sample of 24 FRBs with scattering timescale measurements or limits. 
We infer the Hubble constant ($H_0$) and the DM distribution of the host galaxies simultaneously by fitting our model to the FRB DM measurements. With current data, our results are in agreement with both high and low redshift measurements of $H_0$, obtained using Cosmic Microwave Background (CMB) and Type Ia supernovae data respectively. We project that with about 200 localized FRBs, we would be in a position to distinguish between the two scenarios at 4$\sigma$ confidence. In addition, the host DM is expected to be related to star formation in the host galaxy and the stellar age of the progenitors. \color{black}We show that young progenitors with an age of less than 1 Myr are consistent with our inferred distribution of host DM at 95 percent confidence. \color{black}
These young sources may be associated with long scatter broadening times and large DM from their source environments. Indeed, we find that scatter broadening times of FRBs are inconsistent with the Milky Way ISM, but at the same time, do not appear to be strongly correlated with the FRBs' redshift or with the SFR or stellar mass of their host galaxies. This suggests that scattering is dominated by the immediate environment of the sources.} 

\notoc
\begin{document}
\maketitle
\newpage

\section{Introduction}
Fast radio bursts (FRBs) are millisecond duration cosmological transients and have so far been observed in the frequency range of 100 MHz- 8 GHz. Till date, over 100 FRBs have been localized to their host galaxies (e.g.  \cite{Connor2024, Gao2024,PABC2025}). An important observable of a FRB is its dispersion measure (DM) which is the line of sight integral of electron density as the radio waves travel through the ionized medium. The observed DM of FRBs are measured very precisely. For typical values of DM of the order of few 100 - 1000 pc cm$^{-3}$, the error is of the order O(1) pc cm$^{-3}$ or less. 
Some of these FRBs have additional data on scattering timescales which captures broadening of the radio pulses. The observed DM is a combination of the intergalactic medium (IGM) or a cosmological component, a galactic and halo contribution from our own galaxy as well as the host galaxy contribution. The IGM DM captures relevant information for cosmological applications. Such applications include determining the baryon contents in the Universe, measuring the Hubble constant and probing the era of reionization etc \citep{McQuinn2014,Eichler2017,ZE2018,LGDWZ2018,KL2019,Macquart2020,Beniamini2021,HR2022,James2022,WZW2022,Zhang2023,Khrykin2024,Shaw2024,Gao2024,Connor2024}. 

 The Hubble tension has been a matter of intense interest over the last few years. The inferred Hubble constant ($H_0$) from high $z$ probes such as the CMB \cite{Planck2020} is apparently at a 4$\sigma$ tension with low $z$ direct measurements of Type Ia supernovae \cite{Riess2019} (see \cite{Verde2019} and references therein for more details).  FRBs provide an independent way to infer the  constant and have already been used to determine $H_0$ \cite{Walters2018,Zhao2020,Macquart2020,Hagstotz2022,Wu2022,Liu2023,Fortunato2023,Connor2024}. The IGM DM provides us with a handle on the distance to FRBs, provided we know the average electron density in the universe. In addition, localization of a FRB host galaxy gives us the object's redshift. The distance vs redshift relation provides us with the measurement of $H_0$. Since the electron density fluctuates along different lines of sight, this measurement is noisy. Therefore, with the currently moderately sized sample of localized FRBs, one cannot statistically distinguish between the high redshift and low redshift $H_0$. However, the number of localized FRBs is increasing rapidly. High precision $H_0$ determination using FRBs should be possible within a few years.  

From the available DM data, one can infer the average host DM, even though the host contribution to the total DM for a given FRB is uncertain. From the point of view of using FRBs as a cosmological tool, the host DM adds noise to the determination of $H_0$ from FRB data. That being said the host DM and its evolution with other host galaxy properties carries vital information for our understanding of FRB sources (see e.g. \cite{AB2025}). Both zoom-in galaxy simulations \cite{Beniamini2021,Mo2023} and observations of Galactic pulsars and magnetars (as discussed in this work) find that host DMs vary as a function of stellar age. Therefore the average age of FRB progenitors can be constrained by extracting the host DM contribution from the data, informing us about FRB formation channels. A younger average source age, would suggest that FRBs track the star formation rate, as expected for magnetars formed via core-collapse supernovae (CCSNe) while older ages hint at alternate FRB source formation channels.

Clearly, the use of FRBs as cosmological tools is coupled with the understanding of their progenitor contributions.
In order to extract the IGM component, one has to marginalize over the apriori uncertain host and halo contributions. The host galaxy DM, along with scattering timescale measurements constrain the FRBs' source environment and the typical age of their progenitors. In this work, we infer the rate of expansion of the Universe or Hubble constant as well as the age of FRB progenitors from a sample of localized FRBs simultaneously, by fitting our model to the available data.
We use a sample of 110 FRBs in this paper (Sec. \ref{sec:sample}). Inferring $H_0$ requires modeling the variation of ${\rm DM_{IGM}}$ across different sightlines and redshifts. This is provided by cosmological simulations. Almost all the works in the literature use the Illustris \cite{IllustrisTNG} simulation. In this work, we use the Cosmic Dawn II \cite{CoDa2020} simulation as our fiducial case (Sec. \ref{sec:dispersion}). We find that both simulations provide similar results using current data. The inferred $H_0$ value is consistent with both low and high $z$ measurements within 1$\sigma$ confidence (Sec. \ref{sec:H0}). We project that with about 200 FRBs, we should be in a position to distinguish the two scenarios at $\gtrsim 4\sigma$ confidence (Sec. \ref{sec:projections}). Though this result, in principle, depends somewhat on the redshift distribution of FRBs, it does not change the quantitative results substantially. Further, by comparing our obtained constraint on host DM to the observed DM of Galactic pulsars as a function of their spindown age, we are able to constrain the age of FRB progenitors assuming the pulsar spindown time is a proxy of stellar age. Assuming a single log-normal distribution of host DMs, \color{black}we are able to  show that young progenitors with an age of less than 1 Myr are consistent with inferred host DMs at a 95 percent confidence interval (Sec. \ref{sec:FRB_age}). Older progenitors may appear inconsistent with the inferred DM at face value but an additional contribution from the source or host halo can not be ruled out. \color{black} In Sec. \ref{sec:scat_time}, we use the scattering timescale of a small sample of FRBs to study their correlation with host DMs as well as any information they may carry about their progenitors. Most of the host galaxies in this sample are starforming and, therefore, we need a bigger sample with more quiescent galaxies to test any potential correlation.
We finish with our conclusions in Sec. \ref{sec:conclusions}.     

\section{Sample of FRBs used in this work}
\label{sec:sample}
The observed dispersion measure of FRBs can be partitioned to its individual components as,
\begin{equation}
  \rm{  DM_{obs}=DM_{halo}+DM_{ISM}+DM_{IGM}+\frac{DM_{\rm host}+DM_{source}}{1+z}}
  \label{eq:total_DM}
\end{equation}
where ${\rm DM_{ISM}}$ and ${\rm DM_{halo}}$ are the contributions due to our own galaxy and its circumgalactic medium. The cosmological component, ${\rm DM_{IGM}}$, is contributed by the large scale structure of the universe. It depends sensitively on the evolution of these structures and, therefore, on the expansion history of the universe. ${\rm DM_{host}}$ and ${\rm DM_{source}}$ are due to the host galaxy of the FRB and the immediate surrounding of the FRB respectively which are redshifted to the observer frame. Currently, the host and halo contribution have the largest level of uncertainty. In this work, we used ${\rm DM_{halo}}=50$ pc cm$^{-3}$ as our fiducial value which is in agreement with recent works \cite{Connor2024} (furthermore it is likely an overestimate and thus a conservative choice in this context) \footnote{However, the halo contribution may be anisotropic according, see \cite{DMGNK2021}.}. \color{black} However, it may become important to model this component accurately along with its directional dependence in the future. \color{black}
We compute the ISM contribution using two available models in the literature, NE2001 \citep{CL2002} and YMW2017 \citep{Yao2017}. We use NE2001 as our fiducial model unless otherwise stated. In order to minimize the source contribution we drop 20121102A and 20190520B  from our sample while using their DM as the observable. However, we include them in our scattering timescale sample (Sec. \ref{sec:scat_time}). These sources are likely young, as evidenced by their large rotation measures \citep{Hilmarsson2021,AnnaThomas2023} and persistent radio emission \citep{Chatterjee2017,Niu2022}. Thereby, by removing these sources we ensure minimal contamination due to source terms, if any. In Table \ref{tab:FRB_sample}, we show our sample of 110 localized FRBs along with their observed redshift, total DM, galactic contribution alongside the references. In Sec.  \ref{app:FRB}, we plot the observed DM of our sample of FRBs after subtracting their ISM and halo contribution and compare them with the expected DM from IGM within our fiducial cosmological model. We explain this further in subsequent sections.

\section{Dispersion measure}
\label{sec:dispersion}

\subsection{${\rm DM_{IGM}}$}
The expression for mean IGM contribution can be written as \citep{Beniamini2021},
\begin{equation}
    \langle {{\rm DM_{IGM}}}(z)\rangle=\frac{3cH_0\Omega_{\rm b0}}{8\pi Gm_{\rm p}}\int_0^z {\rm d}z' \frac{(1+z)\xi_e(z)}{\left[\Omega_{\rm m0}(1+z')^3+\Omega_{\Lambda 0}\right]^{1/2}}
    \label{eq:DM_IGM}
\end{equation}
where $\langle {{\rm DM_{IGM}}}(z)\rangle$ is the average contribution and is a function of $z$ (for a given set of cosmological parameters), averaged over different line of sight directions. The parameters $\Omega_{\rm b0}, \Omega_{\rm m0}$ and $\Omega_{\Lambda0}$ are the fractional energy density of baryons, matter and dark energy compared to the critical energy density today which capture the expansion history of the universe. In a flat $\Lambda$CDM universe, $\Omega_{\Lambda0}=1-\Omega_{\rm m0}$. We use the best fit value of these parameters as inferred from the CMB (Cosmic Microwave Background) experiments \citep{Planck2020}. The dispersion measure is a function of the free electron abundance which is captured by $\xi_e(z)$, which in turn depends upon the helium mass fraction $(Y)$. At $z\lesssim 3$, hydrogen and helium is fully ionized \citep{Planck2020,Mcquinn2016}. Therefore, using $Y\approx 0.25$, we have $\xi_e(z)\approx 0.87$.

From Eq. \ref{eq:DM_IGM}, we find that $\langle {{\rm DM_{IGM}}}(z)\rangle \propto \Omega_{\rm b0}H_0$. The degeneracy between $\Omega_{\rm b0}$ and $H_0$ can be broken by using independent measurement such as from CMB (Cosmic Microwave Background) experiments \citep{Planck2020}. CMB primary anisotropy measurement constrain $\Omega_{\rm b0}h^2$ very precisely where $H_0=100h$ kms$^{-1}$Mpc$^{-1}$. From {\it Planck} measurements, we have, $\Omega_{\rm b0}h^2=0.02242\pm 0.00014$ \citep{Planck2020}. For the current expected level of precision from FRB data, we can use the mean value of $\Omega_{\rm b0}h^2$ to break this degeneracy. After fixing this parameter combination, ${\rm DM_{IGM}}$ is inversely proportional to $h$ \citep{James2022}. We take our fiducial value of $h$ to be 0.7 \color{black} and expand the expression for $\langle {{\rm DM_{IGM}}}(z)\rangle$  around this value \color{black}. Therefore, the expression for mean IGM DM at a general $h$ can be parameterized as,
\begin{equation}
    \langle {{\rm DM_{IGM}}}(z,h)\rangle=\langle {{\rm DM_{IGM}}}(z,h=0.7)\rangle \left(\frac{0.7}{h}\right)
    \label{eq:DM_IGM_h}
\end{equation}
The dispersion measure can have a significant scatter along different lines of sight due to fluctuation in matter density. This leads to a variance in the IGM contribution. This quantity is difficult to compute analytically and, therefore, cosmological simulations are used to infer this quantity. As can be expected, this variance can have a sensitive dependence on simulations due to their treatment of baryonic physics. We consider two such simulations below. 
\subsubsection{Cosmic Dawn (CoDa)}
\label{subsubsec:coda}
We choose our fiducial choice of cosmological simulation to be Coda II \citep{CoDa2020} which is a fully-coupled, large-scale, high-resolution, radiation-hydrodynamical
simulation including galaxy formation and reionization up to redshift $z\approx 5.8$. At lower redshifts, the universe is ionized, therefore, the ionization field mimics the matter field. The authors in \cite{Ziegler2024} do a dark matter only, N-body simulation (called "Coda II Dark matter" simulation) to extend the results to $z=0$. These simulations are run on same initial conditions as Coda II. The authors parameterize the standard deviation of ${\rm DM_{IGM}}(z)$. For $z\lesssim 3$. the IGM dispersion measure can be parameterized as a lognormal distribution with mean as in Eq. \ref{eq:DM_IGM} with $\sigma_{\rm IGM}(z)$ as,
\begin{equation}
    \sigma_{\rm IGM}(z)={\rm sinh^{-1}}(0.316z^{-0.677}).
    \label{eq:sigma_IGM}
\end{equation}
We assume this relation to be valid for our fiducial case with $h=0.7$. For different $h$, one needs to redo the simulations. 
However, since $\sigma_{\rm IGM}(z)$ is the variance of a lognormal distribution of ${\rm DM_{IGM}}$ (or $\frac{\rm DM_{IGM}}{\rm \langle DM_{IGM}\rangle }$), we do not expect a strong $h$ dependence. Therefore, we use Eq. \ref{eq:sigma_IGM} as the expression for $\sigma_{\rm IGM}(z)$ without having any dependence on $h$.


\subsubsection{IllustrisTNG}
\label{subsubsec:illustris}
We use IllustrisTNG \citep{IllustrisTNG} as our alternate choice of simulation. Similar to CoDa, it is a large scale gravity+magnetohydrodynamical simulation. The TNG suites include updates to galaxy formation models as well as other technical details which the reader can find in the reference. The host galaxy DM distribution is fitted with the expression \citep{Macquart2020,Zhang2021},
\begin{equation}
    \frac{{\rm d}p_{\rm IGM}}{{\rm d DM_{IGM}}}=A\Delta^{-{\Tilde{\beta}}}{\rm exp}\left[-\frac{(\Delta^{-{\Tilde{\alpha}}}-C_0)^2}{2{\Tilde{\alpha}}^2\sigma^2}\right]
\end{equation}
where $\Delta=\frac{{\rm DM_{IGM}}}{{\rm\langle DM_{IGM}\rangle}}$, ${\Tilde{\alpha}}={\Tilde{\beta}}=3$ and the fitting values of $A, C_0$ and $\sigma$ are provided in \citep{Zhang2021}. Converting the variable from ${\rm DM_{IGM}}$ to ${\rm log(DM_{IGM})}$, we have the equivalent expression,
\begin{equation}
   \frac{{\rm d}p_{\rm IGM}}{{\rm dlog(DM_{IGM})}}=A\times {\rm DM_{IGM}}\times\Delta^{-{\Tilde{\beta}}}{\rm exp}\left[-\frac{(\Delta^{-{\Tilde{\alpha}}}-C_0)^2}{2{\Tilde{\alpha}}^2\sigma^2}\right],
   \label{eq:Illustris_prob}
\end{equation}
The factor $\Delta$ has the dependence on ${\rm\langle DM_{IGM}\rangle}$ which makes it a variable in terms of $h$ using Eq. \ref{eq:DM_IGM_h}. We note that the fitting function $A, C_0$ and $\sigma$ can have a dependence on $h$ as well but we need to simulate these cosmologies in order to get their correct values. In this work, we take the simplifying assumption of these fitting function to be independent of $h$.

\subsection{${\rm DM_{host}}$}
\label{sec:DMhost}
We initially assume the host contribution to be lognormal as is standard in the literature \citep{Macquart2020} and consistent with numerical simulations \cite{Mo2023}. We denote the mean value of the lognormal distribution as $\mu$ . The variance is denoted as $\sigma_{\rm host}$. We assume the fiducial value of $\sigma_{\rm host}$=1 which is roughly the average value obtained from simulations (Table 2 of \cite{Mo2023}).  Additionally, ${\rm DM_{host}}$ (in the rest frame) may be $z$ dependent. This can be parameterized as ${\rm DM_{host}}={\rm DM_{ host,0}}(1+z)^{\alpha}$. As an example, the authors of \cite{Zhang2020} reported an approximate value of $\alpha\approx 1$, on average across all galaxy types, using IllustrisTNG simulation\citep{Nelson2019}. However, current data can only exclude the scenarios with $|\alpha|\gtrsim 2$ at 68 percent confidence \citep{AB2025}. Therefore, in this work, we assume that host galaxy contribution does not evolve with redshift and $\alpha=0$. We denote $\mu_0\equiv\mu(z=0)$, since our canonical model assumes $\alpha=0$, $\mu$ can be replaced with $\mu_0$ in the rest of the paper.

\section{Likelihood evaluation}
\label{sec:Likelihood}
We use Gaussian likelihood for the data analysis. The likelihood of a given FRB with total observed DM (${\rm DM_{i,obs}}$) is given by,
\begin{equation}
\mathcal{L}_i({\rm DM_i'}|z_i)=\int_0^{\rm DM_i'}\frac{{\rm  d}p_{\rm host}({\rm DM_{host}}|\mu_0,\sigma_{\rm host})}{{\rm d DM_{host}}}\frac{{\rm d}p_{\rm IGM}(h)}{{\rm dlog} X}{\rm dDM_{host}},
\label{eq:likelihood}
\end{equation}
where, ${\rm DM_i'=DM_{i,obs}-DM_{i,ISM}-DM_{halo}}$ and $X={\rm DM_i'}-\frac{{\rm DM_{host}}}{1+z_i}$. The distribution of host galaxy DMs is given by,
\begin{equation}
    \frac{{\rm d}p_{\rm host}}{{\rm dDM_{host}}}=\frac{1}{\sqrt{2\pi \sigma^2_{\rm host}}{\rm DM_{host}}}{\rm exp}\left(\rm -\frac{(log(DM_{host})-\mu_0)^2}{2\sigma^2_{\rm host}}\right),
\end{equation}
For results related to CoDa II simulation, we have the IGM distribution as, 
\begin{equation}
    \frac{{\rm d}p_{\rm IGM}(h)}{{\rm dlog} X}=\frac{1}{\sqrt{2\pi \sigma^2_{\rm IGM}}}{\rm exp}\left(\rm -\frac{[{\rm logX}-log(\langle DM_{IGM}\rangle)]^2}{2\sigma^2_{\rm IGM}}\right),
    \label{eq:IGM_prob}
\end{equation}
For results using IllustrisTNG simulation, we use Eq. \ref{eq:Illustris_prob} as the IGM probability distribution while replacing ${\rm DM_{IGM}}$ by X.

The computed likelihood is a function of two parameters $h$ and $\mu_0$. The $h$ dependence is included in $\langle {{\rm DM_{IGM}}}(z,h)\rangle$. Since all FRB sources are independent, the joint likelihood is given by the individual likelihood products,
\begin{equation}
    \mathcal{L}_{\rm tot}=\prod_i^N \mathcal{L}_i,
\end{equation}
where $N$ is the number of FRBs in the sample which is 110 in our case. We use MCMC (Markov Chain Monte Carlo) sampling to scan over the parameter space. 

\section{Dependence of $H_0$ determination on IGM and ISM modeling}
\label{sec:H0}

\begin{figure}[!htp]
\begin{subfigure}[b]{0.4\textwidth}
\includegraphics[scale=0.55]{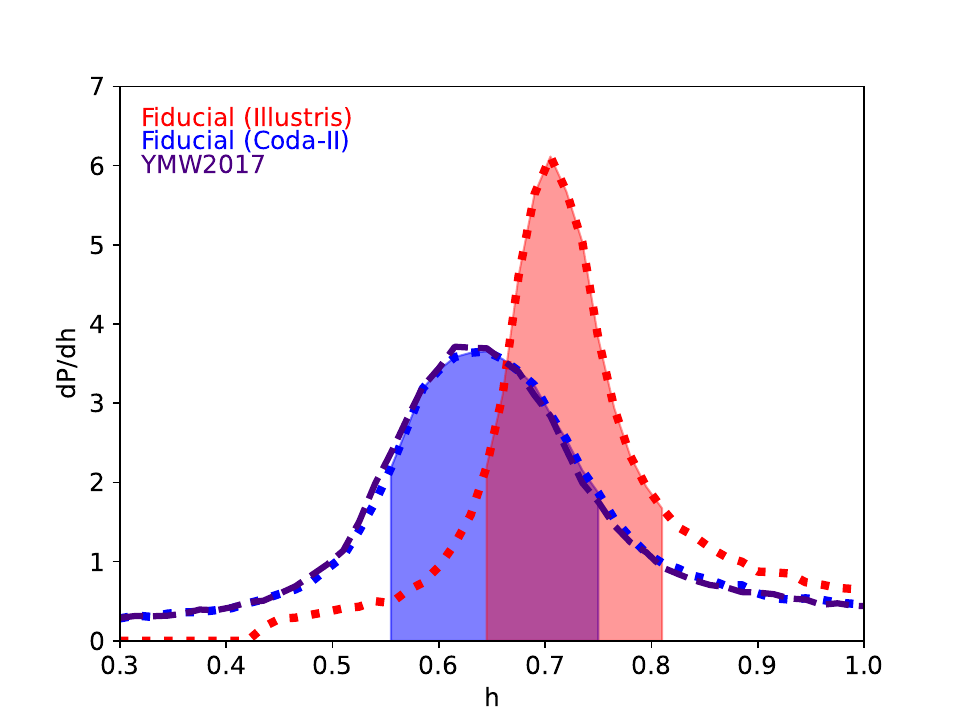}
\end{subfigure}\hspace{50 pt}
\begin{subfigure}[b]{0.4\textwidth}
\includegraphics[scale=0.55]{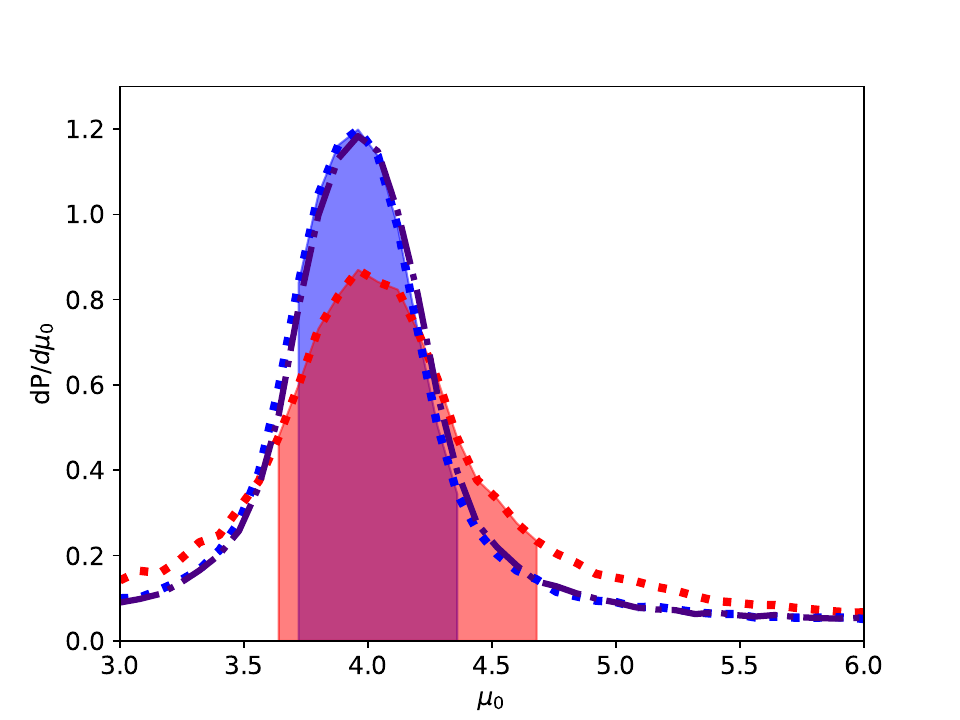}
\end{subfigure}
\caption{Marginalized probability distribution function for $h$ and $\mu_0$ by fitting these parameters to our sample of 110 FRBs. We show the 68 percent confidence interval in shaded regions for respective cases with the same color.  }
 \label{fig:hubble_fiducial}
\end{figure}

We obtain marginalized constraints on $h$ and $\mu_0$ using the results of the two simulations that we have described before. We plot the constraints in Fig. \ref{fig:hubble_fiducial}. The constraint on $h$ is $0.7^{+0.11}_{-0.07}$ and $0.64^{0.11}_{-0.09}$ for IllustrisTNG and CoDa simulations respectively. They are consistent with each other as well with the result of \cite{Connor2024} to within a 1$\sigma$ confidence interval. The best fit value of $\mu_0\approx 4$ applies to both cases.
While our fiducial model above assumed NE2001 as the ISM model, we redo the same computation with the YMW2017 model. We obtain almost identical results. Therefore, we do not expect the choice of ISM model to affect the constraint in $h$.

\section{Future projections on constraining $h$ and viability of resolving Hubble tension}
\label{sec:projections}
Our constraint on $h$ is consistent with respect to high $z$ probes such as the CMB \citep{Planck2020} and low $z$ probes such as Type Ia supernova \citep{Riess2019} within 1$\sigma$ confidence. The Hubble constant inferred from CMB observations ($H_0=67.4^{+0.5}_{-0.5}$ km s$^{-1}$Mpc$^{-1}$) is at $\approx 4\sigma$ tension with low redshift measurements ($H_0=74.0^{+1.4}_{-1.4}$ km s$^{-1}$ Mpc$^{-1}$). It is natural to ask if a larger FRB sample than what is available right now can distinguish between the two $H_0$ values at a high confidence interval. For this purpose, we do a simulation by generating a sample of FRBs with a redshift and ${\rm DM_{ host}}$ distribution. We consider a universe with $h=0.74$ which is what is measured from low redshift probes. 
          
\begin{figure}[!htp]
\centering
 \includegraphics[width=0.7\columnwidth]{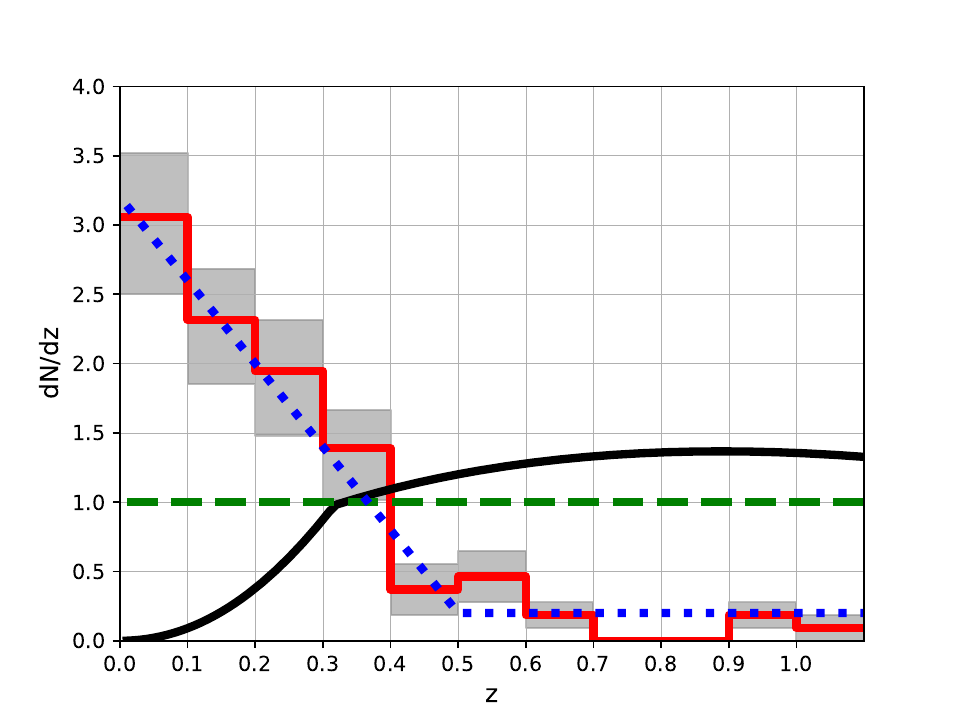}
\caption{Normalized redshift distribution of the FRB sample. In red, we show the number of FRBs in Table \ref{tab:FRB_sample} within $\Delta z=0.1$ intervals along with the Poisson error in gray. The corresponding approximate distribution is shown in dotted blue. We also consider a uniform distribution (dashed green) between $z=0-1$ and a more physically motivated distribution for a hypothetical future large FoV telescope with good localization capabilities (in black; see \S \ref{sec:projections}).}
 \label{fig:redshift_distribution}
\end{figure}

In Fig. \ref{fig:redshift_distribution}, we plot the number of FRBs with known $z$ (from the sample in Table \ref{tab:FRB_sample}) in redshift intervals of 0.1. We use an approximate normalized probability density function in Fig. \ref{fig:redshift_distribution} such that $\int_0^1 {\rm d}z \frac{{\rm dP}}{{\rm d}z}=1$.  The rest frame host DM distribution is assumed to be lognormal with $\mu_0=4$ (best fit value in Fig. \ref{fig:hubble_fiducial}) and $\sigma_{\rm host}=1$ (see \S \ref{sec:DMhost}). We sample from a lognormal distribution of ${\rm DM_{IGM}}$ (with Eq. \ref{eq:DM_IGM_h}, \ref{eq:sigma_IGM} as the mean and variance respectively). At the sample creation stage, we assume the mean value of ${\rm DM_{halo}}=50$ pc cm$^{-3}$ with a uniformly distributed scatter of up to $\pm 30$ pc cm$^{-3}$. At the inference stage, we assume a constant ${\rm DM_{halo}}=50$ pc cm$^{-3}$. For ISM contribution, we assume FRBs to be isotropically distributed in the sky but with positive declination angle. This is due to the sparsity of FRBs with negative declination angle in the current sample. For a given line of sight, we take the average of ${\rm DM_{ISM}}$ of the nearest 4 FRBs which encloses the simulated FRB in terms of sky location. We use the NE2001 model in the sample creation stage and use YMW2017 model in the inference stage. This procedure captures the uncertainty introduced due to our modeling choices.  

\begin{figure}
 \centering
 \includegraphics[width=0.8\columnwidth]{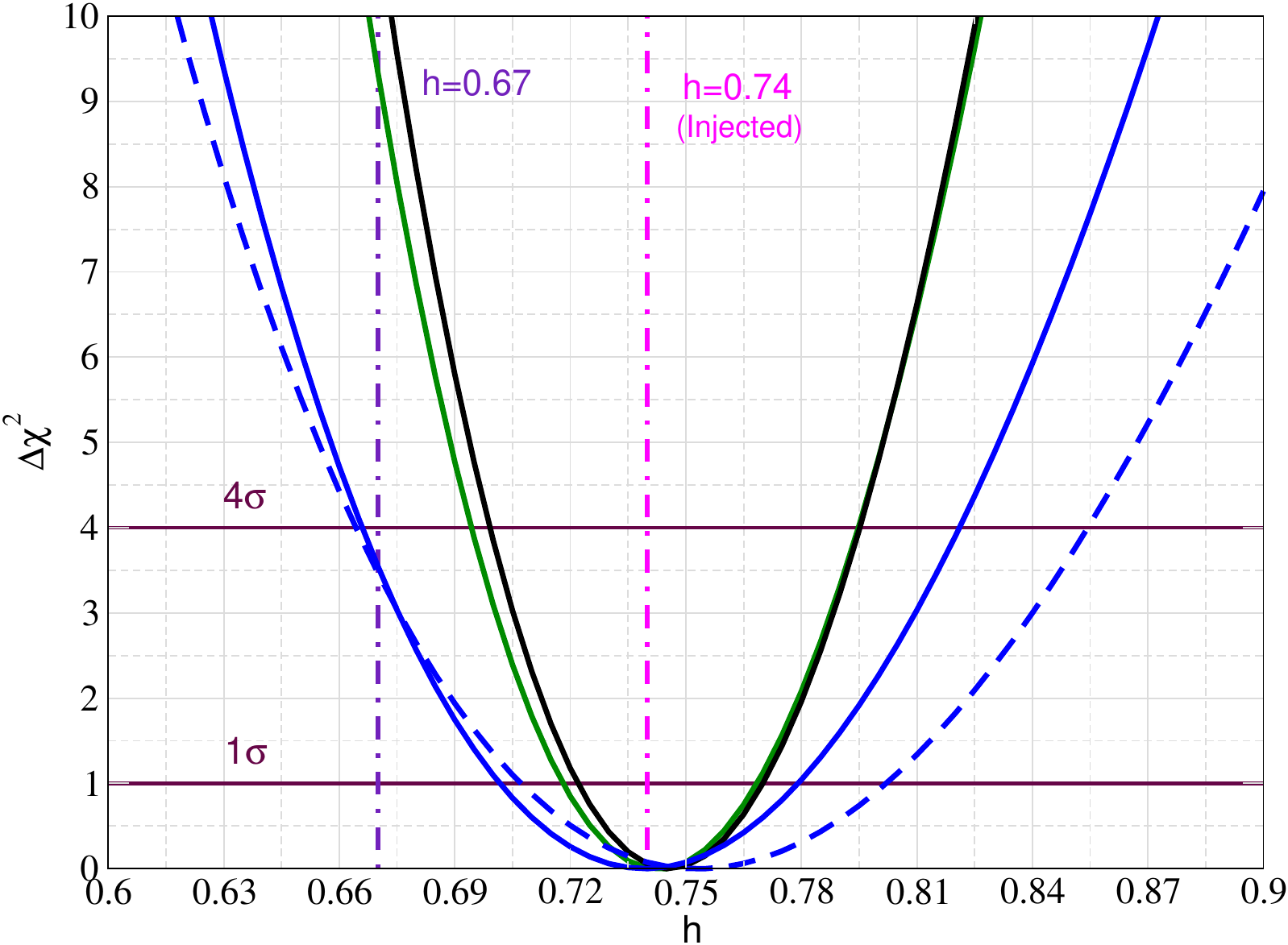}
 \caption{Futuristic constraints from an FRB sample of a given size (denoted in the plot). The color coding follows from Fig. \ref{fig:redshift_distribution}. In dashed blue, we consider a scenario where half of total population of FRBs is contributed by persistent emission-like sources with $\langle {\rm DM}_{\rm host}\rangle \approx 350\mbox{pc cm}^{-3}$ (see Sec. \ref{sec:FRB_age} for details). In our simulation, we create a FRB sample with a background Hubble constant $h=0.74$. We use 200 FRBs for our simulation. With about 200 localized FRBs, we can resolve the Hubble tension at 4$\sigma$ while more number of localized FRBs at higher redshifts will increase the detection significance.  }
 \label{fig:H0_simulation}
\end{figure}

We compute the likelihood of our simulated sample using the same formalism as in Sec. \ref{sec:Likelihood}. We note that the likelihood is a function of two parameters $h$ and $\mu$. We marginalize over $\mu$ to compute the likelihood involving only $h$. We convert it into $\chi^2$ distribution using the definition, $\chi^2(h)=-2\times{\rm log}\mathcal{L}(h)$ where $\mathcal{L}(h)$ is the marginal likelihood over $h$. The absolute value of $\chi^2$ is not physical. Therefore, We compute $\Delta\chi^2(h)$ by taking the minimum value of $\chi^2(h)$ distribution as zero. This quantity is plotted for a few cases in Fig. \ref{fig:H0_simulation}. We find that with our fiducial approximate distribution, $\sim 200$ FRBs would be enough to distinguish between the $H_0$ values reported by low and high redshift probes at 4$\sigma$ confidence. The constraint is expected to be sensitive to the redshift distribution of FRBs. At high redshifts, the observed DM of FRBs is dominated by the IGM component and not polluted by the host contribution. Therefore, we expect to obtain stronger constraint on $h$ compared to our fiducial case, assuming $\Lambda$CDM cosmological model. We simulate such a scenario with a uniform distribution of FRBs over redshift range of 0 and 1.
We also consider a more realistic case where the number of FRBs track the cosmic star formation rate (see \cite{ME2018,Beniamini2021} for more details) and it is assumed that all FRBs are detected above a critical fluence threshold of 1 Jy ms at 0.5 GHz. Other relevant parameters can be found in Sec 2 of \cite{Beniamini2021}. This calculation is representative of a hypothetical survey with good enough localization, such that a large fraction of detected FRBs can also be assigned a redshift (this is as opposed to the current sample of FRBs with $z$ which is clearly biased towards low $z$ - however note that planned surveys such as DSA 2000 \cite{DSA} and CHORD \cite{CHORD} are expected to have $\sim$arcsec localization capabilities or better).
For either of these redshift distributions, we find that we obtain stronger constraints than our fiducial case with 200 FRBs. In either case, we expect to distinguish between the two $H_0$ values with few hundred localized FRBs.

\section{Constraining the age of FRB progenitors}
\label{sec:FRB_age}
\begin{figure}[!htp]
\begin{subfigure}[b]{0.4\textwidth}
\includegraphics[scale=0.57]{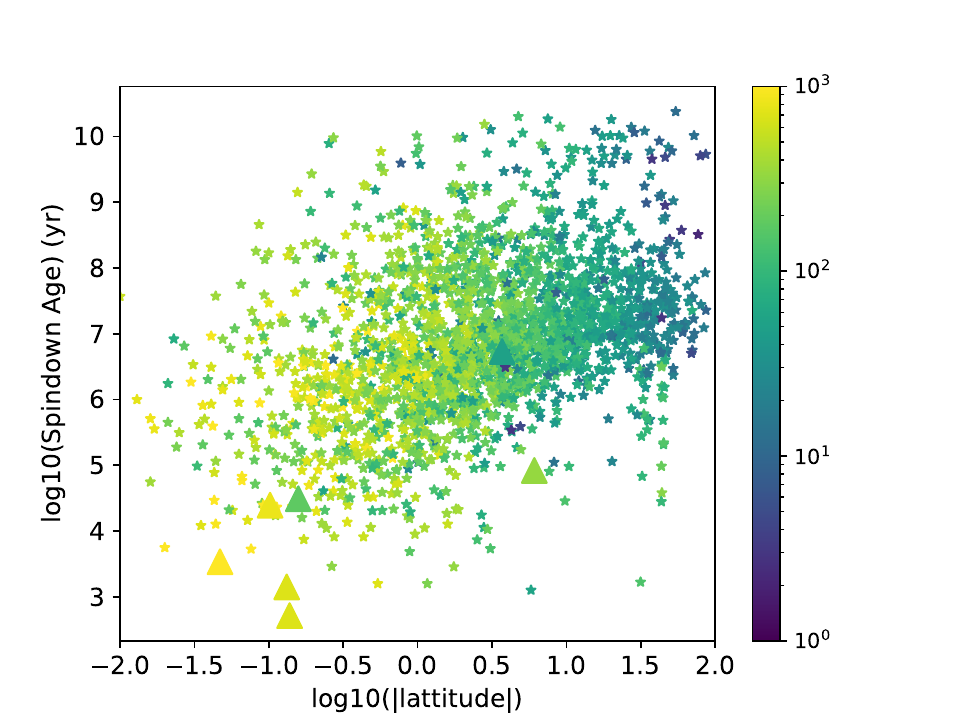}
\end{subfigure}\hspace{70 pt}
\begin{subfigure}[b]{0.4\textwidth}
\includegraphics[scale=0.3]{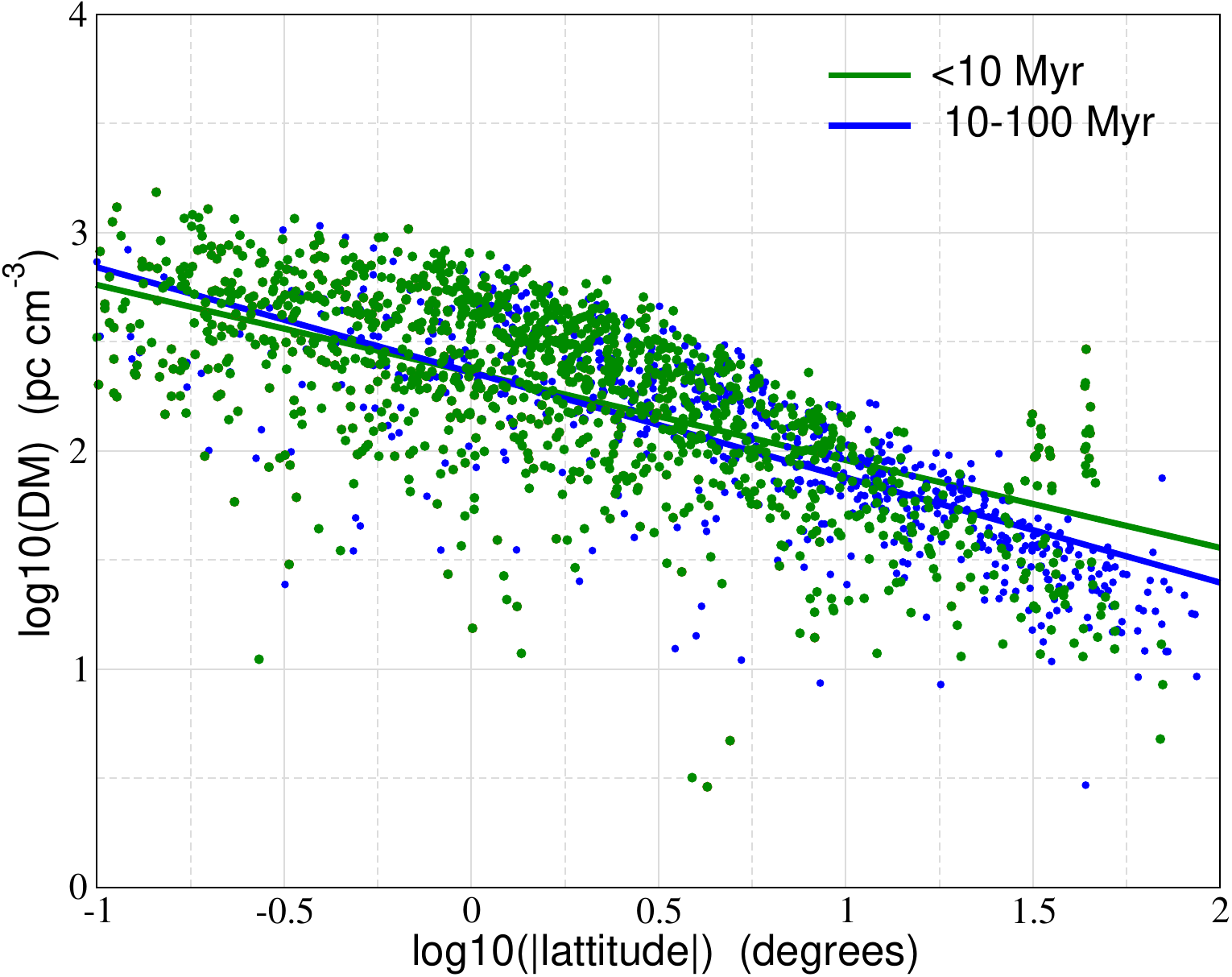}
\end{subfigure}
\caption{We plot the distribution of spindown ages as a function of the absolute value of latitude of the galactic pulsars, color coded in their measured DM (left panel). We have not included millisecond pulsars in this sample. We also show the Galactic radio magnetars in large triangles. These magnetars are not included in any of the analysis. (Right panel) We show the 2D distribution of DM and latitude for pulsars with spindown age $<10$ Myr (green) and 10-100 Myr (blue). We also show the linear best fit for the two cases in the log10(DM) and ${\rm \log_{10}(|latitude|)}$ space.  }
 \label{fig:pulsar_galactic}
\end{figure}

Our inferred ${\rm DM_{host}}$ (Right panel of Fig. \ref{fig:hubble_fiducial}) have implications regarding the age of FRB progenitors. For young progenitors, which are likely to reside in regions of active star formation, we expect higher ${\rm DM_{host}}$ and the opposite at older ages. Our plan is to empirically relate stellar age with ${\rm DM_{host}}$,  using the catalog of pulsars in our own galaxy \footnote{https://www.atnf.csiro.au/research/pulsar/psrcat/}. We use their tabulated spindown age as a proxy for stellar age. \color{black} We plot the distribution of spindown ages of the galactic pulsars as a function of their Galactic latitude ($b$) with their respective DMs in color (Fig. \ref{fig:pulsar_galactic}).
We have excluded the millisecond pulsars in this plot, which are typically recycled and therefore their spindown ages are bad proxies of their true ages.
As the distribution is roughly symmetric between positive and negative $b$, we use the absolute value of $b$ in the plot and further analysis. We also show the Galactic radio magnetars (see \S \ref{sec:Galmagnetars}) in large triangular symbols in Fig. \ref{fig:pulsar_galactic}. We see that the younger systems have, on average, higher DMs and are more likely to be found at lower $b$. This is expected since star formation increases significantly towards the Galactic plane, and stars further from the plane are mostly older systems that have had time to oscillate in the Galactic potential. The average DM decreases at higher $b$, as the column density through the plane probes much more ISM material than towards off-plane latitudes. This can be seen in the right panel of Fig. \ref{fig:pulsar_galactic} where we plot the distribution of pulsars in DM vs $b$ for a given range of spindown ages and show that they are anti-correlated. Moreover, this figure demonstrates that for the range of typical pulsar ages, $b$ is the main factor determining DM (i.e. the age depends on DM mostly because old stars are further off the plane, while the immediate environment of the pulsar is a smaller contributor).
Without taking into account this anti-correlation, the constraints on typical age of FRB host galaxies would be biased. 

As a useful first approximation, one can consider the situation in which a typical FRB host galaxy will have random (isotropically distributed) orientation of its galactic plane relative to our viewing angle  (see \cite{Bhardwajselection,Glowacki2025} for differing views on the potential diversions of FRB hosts from such a distribution and its possible connection to scattering). The orientation angle is denoted as $i$ and $i=\frac{\pi}{2}-b$. For random isotropic orientations, we are expected to have a uniform distribution over ${\rm cos}(i)$. We expect to find fewer galaxies which are edge-on, or $i\sim \pi/2$ ($b\sim 0)$, and therefore that have higher associated host DMs. More galaxies are expected to be found at higher $b$ which, on average, have a lower DM for a given stellar age. In order to obtain the expected host DM distribution for a given stellar age, we sample from a uniform distribution of ${\rm cos}(i)$ and obtain the expected average DM by computing a linear best fit in the ${\rm log(DM)-\log(|b|)}$ space as shown in the right panel of Fig. \ref{fig:pulsar_galactic}. To obtain a smooth DM distribution, we add lognormal scatter with a variance of 0.2 around the best fit line. This variance also roughly accounts for the scatter seen in the right panel of Fig. \ref{fig:pulsar_galactic}. 
\color{black}

We also take into account the mismatch of stellar mass of our galaxy and the typical FRB host galaxy. We note that the stellar mass of our galaxy is $\approx 5\times 10^{10} M_{\odot}$ \cite{HPCFL2007} while a typical FRB host galaxy has a stellar mass of the order of $10^{10} M_{\odot}$ (Fig. 2 of \cite{Sharma2024}). Since, the DM distribution is a function of stellar mass of the host galaxy, we correct for this mismatch. To this end, we approximate the results from \cite{Beniamini2021} which used the FIRE simulation \cite{Ma2018} to obtain the host DM distribution as a function of stellar mass at high redshifts ($z\gtrsim 5$) (see Fig. 6 of \cite{Beniamini2021}). The authors show the DM distribution for a few different stellar mass in the range of $10^4-10^{10} M_{\odot}$. The main effect of changing the galaxy stellar mass in these simulations is to shift the peak of the DM$_{\rm host}$ distribution, such that approximately $\mbox{DM}_{\rm host} \propto M_*^{0.3}$. We use this to scale up the host DM by a factor of 2 from typical FRB host galaxy mass $10^{10} M_{\odot}$ 
 to our galaxy with a stellar mass $5\times 10^{10} M_{\odot}$. 
 A similar dependence of the host DM on stellar mass is seen in Fig. 5 of \cite{Mo2023} which uses IllustrisTNG (though the results can depend upon the details of the simulation).   

\begin{figure}
\centering
\includegraphics[width=0.7\columnwidth]{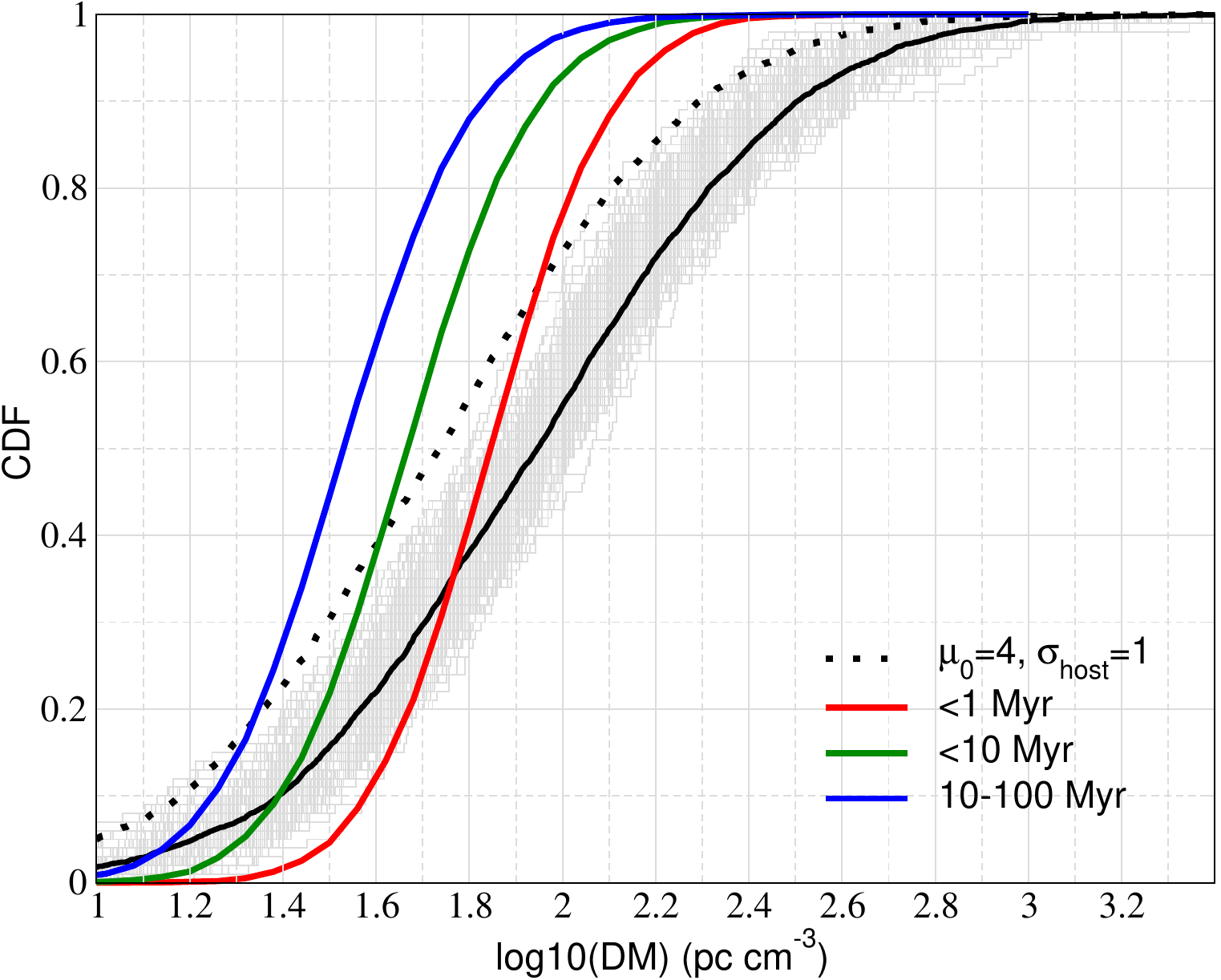}
\caption{Comparison of cumulative distribution function (CDF) of our inferred host DM contribution with the DM \color{black}distribution of Galactic pulsars sampled as a function of viewing angle and  as a function of age (See text for details). \color{black} In dashed and solid black, we consider our inferred best fit value of $\mu_0=4$ and the rescaled version respectively. We consider a sample of 5000 FRBs to obtain a smooth curve.  We also plot realizations where we  sample 100 FRBs from a lognormal distribution with $\mu_0=4$ and $\sigma_{\rm host}=1$, in grey. 
}
 \label{fig:host_cont}
\end{figure}

We plot the cumulative distribution function (CDF) of the expected DM of host galaxies, obtained by the above procedure, for a couple of cases with prescribed ages in Fig. \ref{fig:host_cont}. We note that for younger ages, the typical DM$_{\rm host}$ is larger. We compare these cases with our inferred results for $\mu_0$. We draw samples of 100 FRBs from a lognormal distribution with mean $\mu_0=4$ and $\sigma_{\rm host}=1$ and plot their CDF (after rescaling), repeating this process many times. With a single log-normal population of DM$_{\rm host}$, \color{black} we find that the scenario in which the FRB progenitors are younger than 1 Myr is consistent with our inferred host DM at a 95 percent confidence interval. The older population have even lower DM but an extra contribution from source or the host halo can not be ruled out. \color{black}




As an additional check of our results, we assume a bimodal distribution of host galaxy DM ($\mu_0$) with an addition of high DM component which is contributed by persistent emission-like sources (e.g. FRB 20121102A and FRB 20190520B, \cite{Niu2022,Jahns2023}) are likely young and high-magnetic field strength magnetars \citep{RABG2025}. This is motivated by the fact that the sample of well studied FRBs clearly includes both sources in highly active star forming regions, with a significant DM component associated with the sources' immediate environment and host galaxies (as described above) and other sources, such as FRB 20200120E, which has been localized to an old environment - a globular cluster in M81, and for which the DM contribution from the source and host is constrained to be $\lesssim 20 \mbox{pc cm}^{-3}$ \cite{Bhardwaj2021,Kirsten2022}.
We assume this large DM component to have $\mu_{0,{\rm high}} \approx 5.86$ ( or ${\rm e}^{\mu_{0,{\rm high}}}\approx 350$ pc cm$^{-3}$) and $\sigma_{\rm host}=1$, consistent with the population of persistent emission-like sources. We assume the large DM population to be a fraction $f$ of total population of FRBs. We fit our sample of 110 FRBs with this bimodal distribution of $\mu_0$ (fitting $\mu_{0,{\rm low}}$)  in addition to varying $h$ and $f$. We find that $f$ is consistent with zero with 1$\sigma$ upper limit to be 0.3. The current data is not powerful enough to constrain $f$ at high significance, and the large DM population may still be the overall majority. Therefore, we do not pursue this issue further in this work. We note that, in this case, the estimate for $h$ becomes $0.66^{+0.28}_{-0.16}$.

\section{Scattering timescale}
\label{sec:scat_time}
\begin{figure}[!htp]
\begin{subfigure}[b]{0.4\textwidth}
\includegraphics[scale=0.29]{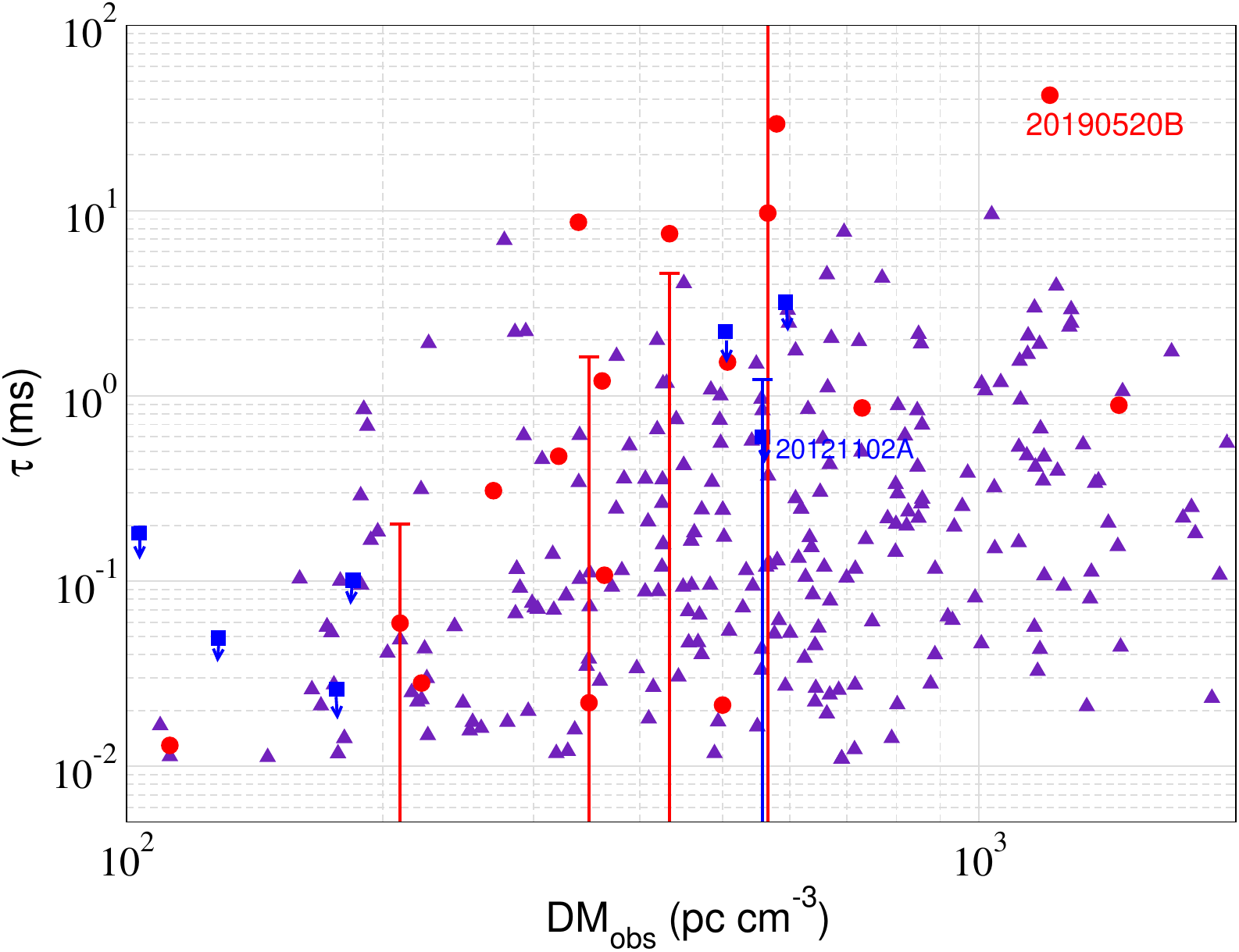}
\end{subfigure}\hspace{50pt}
\begin{subfigure}[b]{0.4\textwidth}
\includegraphics[scale=0.29]{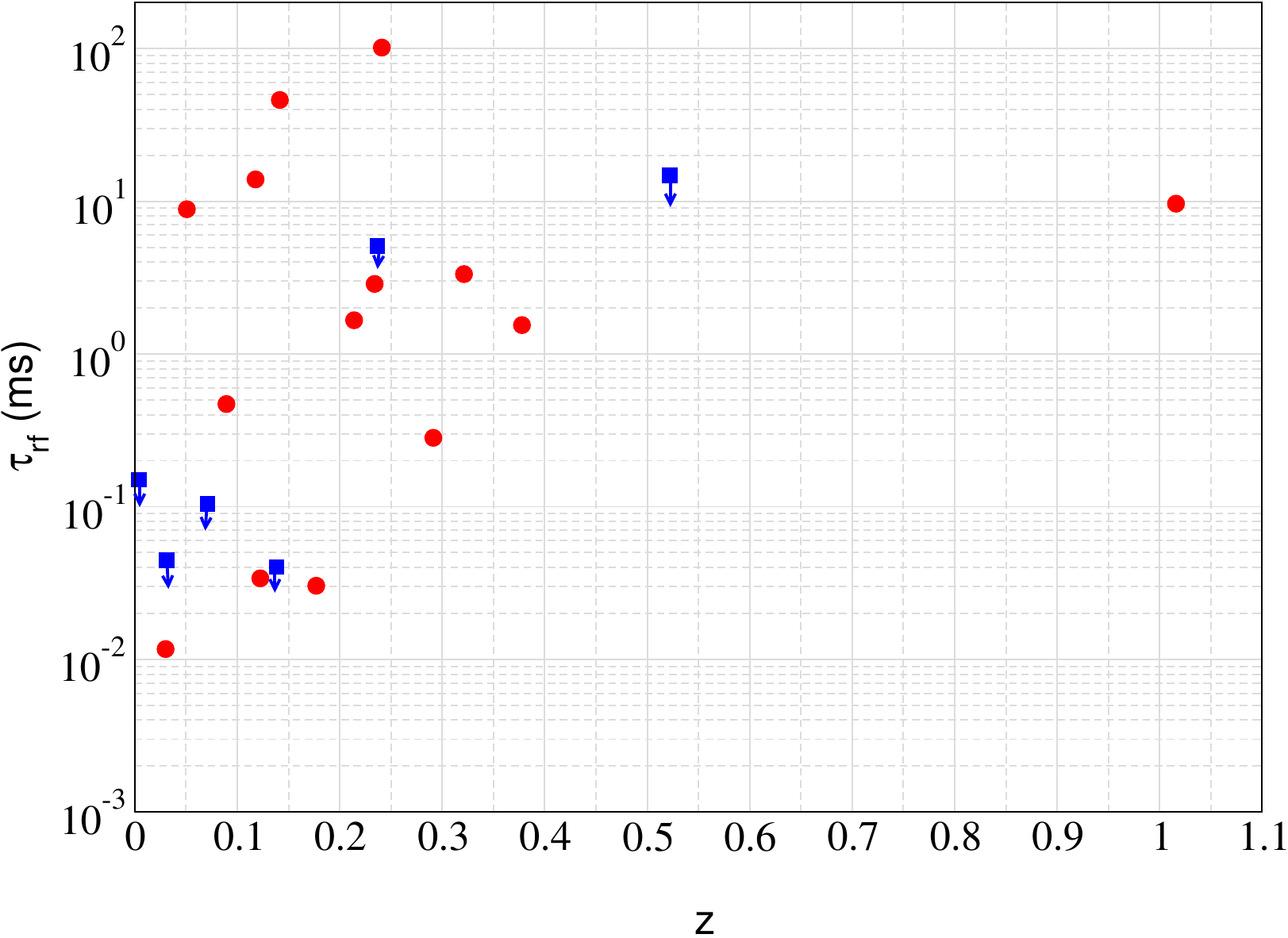}
\end{subfigure}\hspace{30 pt}
\begin{subfigure}[b]{0.4\textwidth}
\includegraphics[scale=0.29]{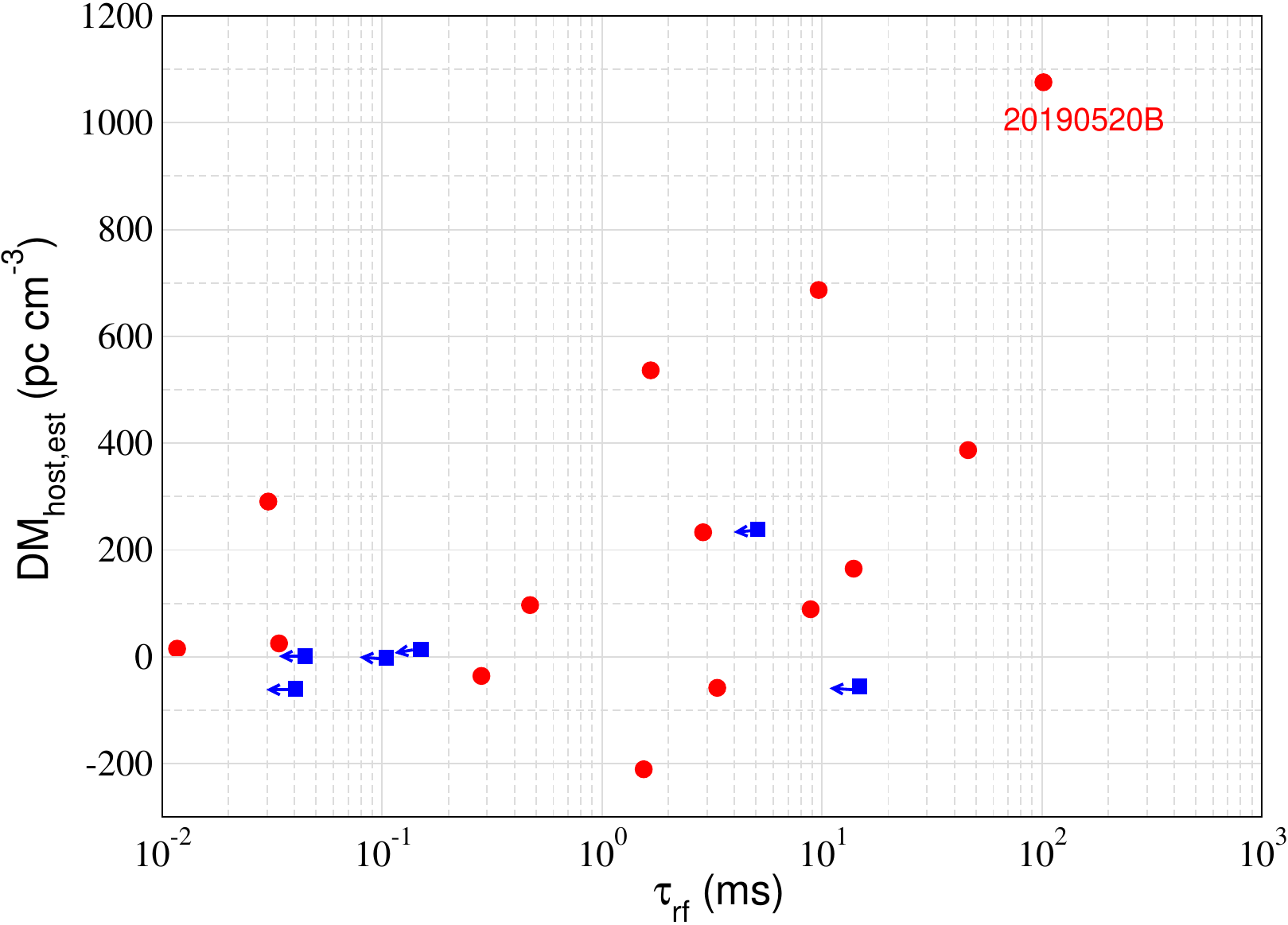}
\end{subfigure}\hspace{46.5 pt}
\begin{subfigure}[b]{0.4\textwidth}
\includegraphics[scale=0.51]{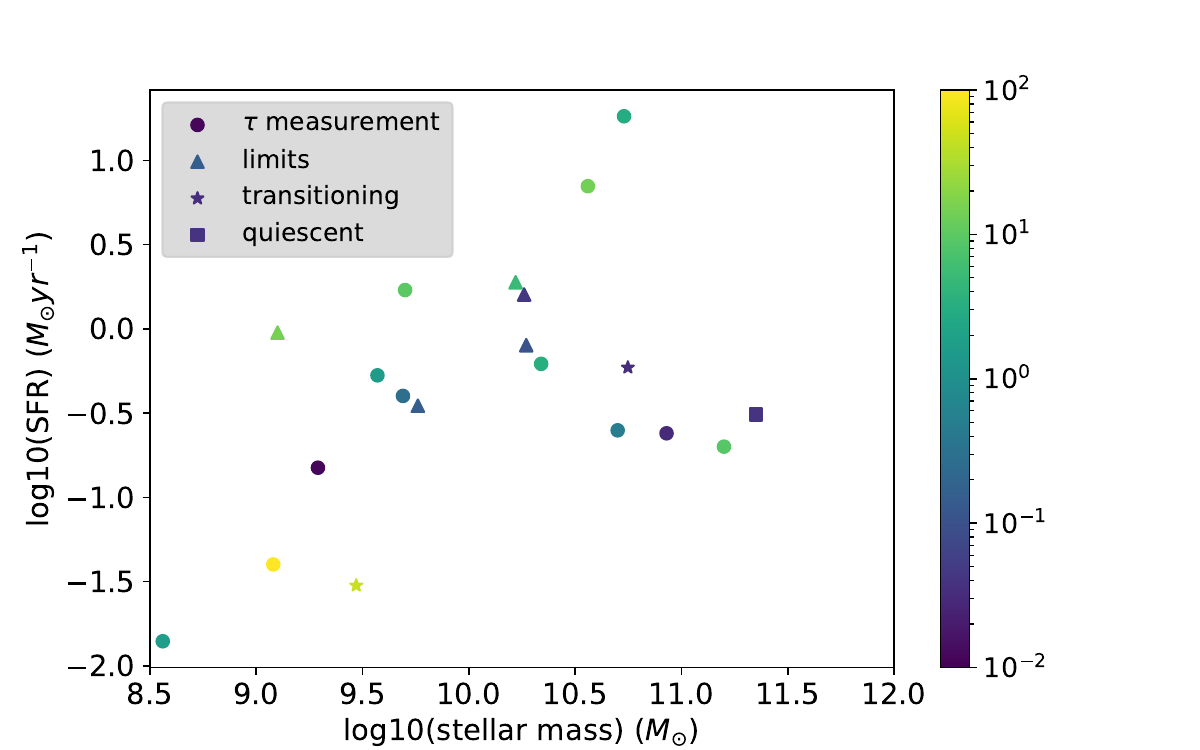}
\end{subfigure}\hspace{40 pt}

\caption{Dependence of $\tau$ as a function of other FRB properties. Upper left: observed $\tau$ vs observed DM. The corresponding Milky Way contribution to $\tau$ is shown with errorbars using the model of \cite{Cordes2016} and is consistent with measured values in 4 FRBs (these are excluded in the following). We also show the CHIME catalogue in triangle points \cite{Chime_catalog}. Upper right: $\tau_{\rm rf}$ (rest frame of host galaxy) vs $z$. Lower left: Our rough estimate of host galaxy DM (see text) as a function of $\tau_{\rm rf}$.  Lower right: $\tau_{\rm rf}$  as a function of stellar mass and star formation rate (circles and triangles denote cases with measured $\tau$ and its upper limit correspondingly). The two transitioning galaxies are denoted by stars. The quiescent galaxy which has a lower limit on $\tau$ is denoted by a square. } 
\label{fig:scattering_plot}
\end{figure}

We consider the correlation of the scattering timescale ($\tau$) of FRBs as a function of redshift and total DM. We tabulate a sample of localized FRBs in Table \ref{tab:scattering_sample} with associated $\tau$ or available upper limits. 
We have taken the observed frequency to be at 1 GHz and have converted available data at different frequencies assuming $\tau \propto \nu^{-4.4}$. We also show the stellar mass and the star formation rate (SFR) of the host galaxy along with galaxy type. In upper left panel of Fig. \ref{fig:scattering_plot}, we plot the observed $\tau$ vs the observed DM. The data is suggestive of a potential correlation between observed $\tau$ and ${\rm DM_{obs}}$. Previously, The authors in \cite{Ravi2019_1} and \cite{GFFBDDL2022} have noted the correlation between $\tau$ and extragalactic component of DM for unlocalized FRBs. As the measured scattering is generally inconsistent with expectations from the IGM \cite{BK2020}, this could suggest an underlying correlation between $\tau$ and ${\rm DM_{host}}$. We also plot the measurements from the CHIME catalog \cite{Chime_catalog} which do not show any significant correlation between $\tau$ and ${\rm DM_{obs}}$. For the sample of localized FRBs studied here, we compute the expected $\tau$ from our galaxy using the fit obtained in \cite{Cordes2016}. We show them in vertical lines with error bars in Fig. \ref{fig:scattering_plot} (top right panel). Only 4 of these FRBs are consistent with Galactic $\tau$. We exclude these 4 FRBs from further analysis.
We define $\tau_{\rm rf}$ as the scattering timescale in the host galaxy rest frame (at 1 GHz). Assuming $\tau\propto \nu^{-4.4}$ (as typically expected for a Kolmogorov spectrum of turbulence, see \cite{BK2020}), $\tau_{\rm rf}$ is related to the observed $\tau$ by the relation, $\tau_{\rm rf}=\tau (1+z)^{3.4}$ where $z$ is the redshift of the host galaxy.
We plot $\tau_{\rm rf}$ as a function of $z$ (upper right panel). We do not find a strong correlation between ${\rm log}\tau_{\rm rf}$ and $z$ (Pearson correlation coefficient $r$= 0.27 and $p$-value=0.35), suggesting that any $\tau$-$\mbox{DM}_{\rm obs}$ correlation is more likely driven the host galaxy (+ source) contribution. 

In the bottom left panel of Fig. \ref{fig:scattering_plot}, we plot our estimate of host galaxy DM as a function of $\tau_{\rm rf}$. Our estimate follows from Eq. \ref{eq:total_DM} with $\rm{DM_{host,est}=  (DM_{obs}-DM_{halo}-DM_{ISM}-\langle DM_{IGM}\rangle )}(1+z)$ where $ {\rm \langle DM_{IGM}\rangle} $ follows from Eq. \ref{eq:DM_IGM}. For some of the objects our estimated host DM is negative which is expected since the IGM component varies along line of sight in contrast to our assumption in the equation above. We see a hint of correlation between $\tau_{\rm rf}$ and $\rm{DM_{host,est}}$ ($r$ coefficient=0.745, $p$-value=0.002). We also plot $\tau_{\rm rf}$ as a function of stellar mass and SFR. Most of the host galaxies are  starforming and they show a significant scatter in their star formation rate, stellar mass, and their ratio (the specific star formation rate). 
There are two transitioning galaxies and one quiescent galaxy in this sample. Overall, with this small sample, there is no significant correlation between $\tau_{\rm rf}$ and these galaxy properties and no clear
difference between the transitioning and starforming galaxies. This may suggest that scatter broadening is dominated by the immediate surrounding environment rather than the host galaxy. However, we need a bigger sample with more quiescent galaxies to test whether the scattering timescale can be a good proxy to test FRB host galaxy environment. 


   \begin{table}[H]
  \begin{center}
   \begin{tabular}{l|c|c|c|c|c|r} 
   FRB & z & ${\rm log10(M_*)}$ $M_{\odot}$ & SFR ($M_{\odot}$ yr$^{-1}$) &  $\tau$ (ms) & galaxy type & References \\
    \hline
    20181030A & 0.0039 & 9.76 & 0.35 & $<0.18$ & starforming & \cite{Chime_catalog} \\
    20181220A & 0.027 & $9.86^{0.14}_{-0.12}$ & $2.9^{1.6}_{-0.9}$ & $0.059\pm 0.004$ & starforming & \cite{Chime_catalog} \\
    20181223C & 0.03 & $9.29^{0.16}_{-0.20}$ & $0.15^{0.12}_{-0.08}$ & $0.013\pm 0.003$ & starforming & \cite{Chime_catalog} \\
    20190425A & 0.031 & $10.26^{0.09}_{-0.1}$ & $1.6^{1.5}_{-0.9}$ & $<0.049$ & starforming & \cite{Chime_catalog} \\
   20180916 & 0.0337 & $9.91_{-0.05}^{0.03}$ & $0.04_{-0.02}^{0.03}$ & 0.022 & starforming & \cite{Marcote2020} \\
    20201123 & 0.0507 & 11.2 & 0.2 & 7.5 & starforming & \cite{Rajwade2022} \\
    20210405I & 0.066 & 11.25 & 0.3 & $9.7\pm 0.2$ & starforming & \cite{Driessen2024} \\
   20190418A & 0.071 & $10.27^{0.13}_{-0.17}$ & $0.8^{1.1}_{-0.6}$ & $<0.1$ & starforming & \cite{Chime_catalog} \\
   20220509 & 0.0894 & $10.7\pm 0.01$ & $0.25_{-0.04}^{0.07}$ & $0.3\pm 0.07$ & starforming & \cite{Connor2023} \\
      20190608 & 0.1178 & $10.56\pm 0.02$ & $7.03_{-1.15}^{1.43}$ & $8.65\pm 0.52$ & starforming & \cite{Day2020} \\
   20190110C & 0.1224 & 10.748 & 0.59 & $0.028\pm 0.004$ &  transitioning & \cite{Chime_catalog} \\ 
   20240209 & 0.1384 & $11.35\pm 0.01$ & $<0.31$ & $<0.026$ & quiescent & \cite{Eftekhari2025,Shah2025} \\
   20210410 & 0.1415 & $9.47\pm 0.05$ & $0.03_{-0.01}^{0.03}$ & $29.4\pm 2.8$ & transitioning & \cite{Caleb2023} \\
   20210603A & 0.177 & $10.93\pm 0.04$ & $0.24\pm 0.06$ & $0.02\pm 0.0004$ & starforming & \cite{Cassanelli2024} \\
   20121102A & 0.193 & $8.14_{-0.10}^{0.09}$ & $0.05_{-0.01}^{0.02}$ & $\lesssim 0.6$ & starforming & \cite{Ocker2021} \\
   20210117A & 0.214 & $8.56_{-0.08}^{0.06}$ & $0.014_{-0.004}^{0.008}$ & 0.86 & starforming & \cite{Bhandari2023} \\
    20191001 & 0.234 & $10.73_{-0.08}^{0.07}$ & $18.28_{-8.95}^{17.24}$ & $1.52\pm 0.92$ & starforming & \cite{Bhandari2020} \\
   20190714 & 0.2365 & $10.22\pm 0.04$ & $1.89_{-0.72}^{1.22}$ & $\lesssim 2.23$ & starforming & \cite{Qiu2020} \\
   \end{tabular}
   \end{center}
   \end{table}
  \begin{table}[H]
  \begin{center}
   \begin{tabular}{l|c|c|c|c|c|r} 
   FRB & z & ${\rm log10(M_*)}$ $M_{\odot}$ & SFR ($M_{\odot}$ yr$^{-1}$) &  $\tau$ (ms) & galaxy type & References \\
   \hline
   20190520B & 0.241 & $9.08_{-0.09}^{0.08}$ & $0.04_{-0.02}^{0.04}$ & $41.99\pm 5.75$ & starforming & \cite{Ocker2023} \\
   20190102 & 0.2913 & $9.69_{-0.11}^{0.09}$ & $0.4_{-0.11}^{0.31}$ & $0.1\pm 0.005$ & starforming & \cite{Day2020} \\
   20180924 & 0.3212 & $10.3\pm 0.02$ & $0.62_{-0.24}^{0.32}$ & $1.2\pm 0.04$ & starforming & \cite{Bannister2019} \\
   20190611 & 0.3778 & $9.57\pm 0.12$ & $0.53_{-0.26}^{0.77}$ & $0.47\pm 0.52$ & starforming & \cite{Day2020} \\
   20190711 & 0.5220 & $9.1_{-0.23}^{0.15}$ & $0.95_{-0.5}^{0.96}$ & $\lesssim 3.2$ & starforming & \cite{Qiu2020} \\
   20220610A & 1.016 & 9.7 & 1.7 & 0.89 & starforming & \cite{Ryder2023} \\
      \end{tabular}
   \caption{List of localized FRBs with measured scattering timescales at 1 GHz, their stellar mass, star formation rate which are used to obtain results in this section. We also provide the error bars in these measurements wherever available.  } 
   \label{tab:scattering_sample}
   \end{center}
   \end{table}

 \label{fig:DMhost_SFR}

\section{Conclusions}
\label{sec:conclusions}
We have tested some cosmological and astrophysical implications of the observed sample of FRBs with known $z$. First, we infer the Hubble constant using a sample of 110 localized FRBs. This requires inputs from cosmological simulations. We have used results from two simulations and show that they give consistent results at present. Our inferred $H_0$ is consistent with both low and high redshift measurements at 1$\sigma$ confidence interval. We project that with $\sim 200$ localized FRBs, we should be able to distinguish between the two measurements at 4$\sigma$ confidence interval. This projection depends upon the redshift distribution of the sample of FRBs with known $z$. At high redshifts, the cosmological component of the DM dominates over the host DM contribution. Therefore, the observed DM is polluted less by the host DM and we have a better chance to distill $H_0$ from the dominant cosmological component. Projecting towards the future, FRBs can provide a unique angle on the Hubble tension debate. The reason is that the measurement of $H_0$ with FRBs can in principle be done over a wide range of redshifts, up to the era of H reionization \cite{Beniamini2021}. This can allow to get a low and high $z$ estimate of $H_0$ using a single technique and to determine whether the Hubble tension is due to measurement systematics or whether it is indicative of new physics.

We have emphasized in this work the importance of simultaneously inferring $H_0$ and $\mu_0$ (the host DM contribution) in order to make the best and most reliable use of FRB data to infer cosmological parameters.
The inferred DM$_{\rm host}$ is typically $\sim 60 \mbox{pc cm}^{-3}$, \color{black} which naively may seem much smaller than young pulsars and magnetars in the Milky Way, for which the measured values are on average at least four times larger. 
However, we show that there is an anti-correlation between DM and latitude with higher DM objects preferentially found at lower latitudes, and lower latitudes being correlated with younger stellar ages for sight-lines within the Milky Way. Accounting for this anti-correlation, we find that young progenitors with ages of less than 1 Myr are consistent with our inferred host DM distribution at a 95$\%$ confidence interval. Older populations may be inconsistent, however, an additional contribution from near the sources or host halos can not be ruled out. These results are consistent with magnetars as the progenitors of all or most of FRBs, and supports independent lines of evidence along those lines (see  e.g. \cite{BK2025a} and references therein).\color{black}  

That being said, our methodology, as applied to the current data, can only obtain broad characteristics of the FRB DM$_{\rm host}$ population and cannot exclude even a tens of percent sub-population with much larger DM$_{\rm host}$, consistent with persistent emission-like sources, which may be associated with particularly young and large magnetic field strength magnetars. An indication of a potentially bimodal of FRB source ages (and DM$_{\rm host}$) is motivated by the contrast between the environments of FRBs such as 20121102A and 20190520B and that of, e.g. FRB 20200120E. In addition, studying the observed population of non-localized FRBs \cite{Gupta2025} have recently shown that such a bimodal population of FRB source ages describes the current data better than a single (star formation following or mass following) population.
Furthermore, as reported in the first CHIME catalogue, CHIME detected bursts have an observational selection effect against detecting highly scattered FRBs, with $\tau\gtrsim 100$\,ms \cite{CHIME2021}. As long scatter broadening times are potentially correlated with larger DM$_{\rm host}$ (see figure \ref{fig:scattering_plot}), the intrinsic population of large DM$_{\rm host}$ FRBs might even be larger than the low DM$_{\rm host}$ population.

Finally, we have considered the correlation between the observed scattering timescale and observed DM. Our sample of FRBs with scattering timescale measurements is dominated by starforming galaxies, with just two transitioning galaxy hosts and one quiescent galaxy. The starforming galaxies show significant variation in their associated specific star formation rate and these broad galaxy characteristics aren't strongly correlated with the scattering timescale. There is a tentative correlation between the scatter broadening time and the DM$_{\rm host}$ estimates, suggesting that scattering may be dominated by the more immediate environments of FRB sources rather than the host galaxy interstellar medium. For a turbulent plasma screen, one expects $\tau_{\rm rf}\propto \nu^{-22/5}d_{\rm scr} \mbox{DM}_{\rm scr}^{12/5}$ (e.g. \cite{2025Natur.637...48N}), where $d_{\rm scr}$ is the distance of the screen from the source and DM$_{\rm scr}$ is the dispersion measure associated with the screen. A correlation between $\tau_{\rm rf}$ and DM$_{\rm host}$ is therefore expected to be most clearly visible in bursts with large $\tau_{\rm rf}$ / excess DM, for which the plasma screen would dominate both. Indeed, in a recent work, \cite{2025arXiv250301740C} have shown that the immediate environment of FRB 20190520B, has a sSFR that is $\gtrsim 3$ times larger the average value within the host. Considering that this source is also the largest in our sample in terms of its $\tau_{\rm rf}$ / excess DM, this is highly suggestive of those properties being dominated by the near environment of the source, rather than the ISM of its host galaxy.
A larger sample of localized FRBs with constraints on scatter broadening and galaxy properties and more in-depth observations of their environments (e.g. constraining H$\alpha$ emission) will be needed to test this picture.

\section*{Acknowledgements}
We thank Pawan Kumar and Om Gupta for helpful discussions.
SKA is supported by the ARCO fellowship. PB is supported by a grant (no. 2020747) from the United States-Israel Binational Science Foundation (BSF), Jerusalem, Israel, by a grant (no. 1649/23) from the Israel Science Foundation and by a grant (no. 80NSSC 24K0770) from the NASA astrophysics theory program.


{
\vspace{-3mm}
\bibliographystyle{unsrtads}
\bibliography{main}

\begin{thebibliography}{100}

\bibitem{Connor2024}
Liam {Connor}, Vikram {Ravi}, Kritti {Sharma}, Stella~Koch {Ocker}, Jakob {Faber}, Gregg {Hallinan}, Charlie {Harnach}, Greg {Hellbourg}, Rick {Hobbs}, David {Hodge}, Mark {Hodges}, Nikita {Kosogorov}, James {Lamb}, Casey {Law}, Paul {Rasmussen}, Myles {Sherman}, Jean {Somalwar}, Sander {Weinreb}, and David {Woody}.
\newblock {A gas rich cosmic web revealed by partitioning the missing baryons}.
\newblock {\em arXiv e-prints}, page arXiv:2409.16952, September 2024.
\newblock \href {http://arxiv.org/abs/2409.16952} {\path{arXiv:2409.16952}}, \href {http://dx.doi.org/10.48550/arXiv.2409.16952} {\path{[DOI]}}, {\small[\href{https://ui.adsabs.harvard.edu/abs/2024arXiv240916952C}{ADS}]}.

\bibitem{Gao2024}
D.~H. {Gao}, Q.~{Wu}, J.~P. {Hu}, S.~X. {Yi}, X.~{Zhou}, and F.~Y. {Wang}.
\newblock {Measuring Hubble constant using localized and unlocalized fast radio bursts}.
\newblock {\em arXiv e-prints}, page arXiv:2410.03994, October 2024.
\newblock \href {http://arxiv.org/abs/2410.03994} {\path{arXiv:2410.03994}}, \href {http://dx.doi.org/10.48550/arXiv.2410.03994} {\path{[DOI]}}, {\small[\href{https://ui.adsabs.harvard.edu/abs/2024arXiv241003994G}{ADS}]}.

\bibitem{PABC2025}
Eduard~Fernando {Piratova-Moreno}, Luz {{\'A}ngela Garc{\'\i}a}, Carlos~A. {Benavides-Gallego}, and Carolina {Cabrera}.
\newblock {Fast Radio Bursts as cosmological proxies: estimating the Hubble constant}.
\newblock {\em arXiv e-prints}, page arXiv:2502.08509, February 2025.
\newblock \href {http://arxiv.org/abs/2502.08509} {\path{arXiv:2502.08509}}, \href {http://dx.doi.org/10.48550/arXiv.2502.08509} {\path{[DOI]}}, {\small[\href{https://ui.adsabs.harvard.edu/abs/2025arXiv250208509P}{ADS}]}.

\bibitem{McQuinn2014}
Matthew {McQuinn}.
\newblock {Locating the ``Missing'' Baryons with Extragalactic Dispersion Measure Estimates}.
\newblock {\em \apjl}, 780(2):L33, January 2014.
\newblock \href {http://arxiv.org/abs/1309.4451} {\path{arXiv:1309.4451}}, \href {http://dx.doi.org/10.1088/2041-8205/780/2/L33} {\path{[DOI]}}, {\small[\href{https://ui.adsabs.harvard.edu/abs/2014ApJ...780L..33M}{ADS}]}.

\bibitem{Eichler2017}
David {Eichler}.
\newblock {Nanolensed Fast Radio Bursts}.
\newblock {\em \apj}, 850(2):159, December 2017.
\newblock \href {http://arxiv.org/abs/1711.04764} {\path{arXiv:1711.04764}}, \href {http://dx.doi.org/10.3847/1538-4357/aa8b70} {\path{[DOI]}}, {\small[\href{https://ui.adsabs.harvard.edu/abs/2017ApJ...850..159E}{ADS}]}.

\bibitem{ZE2018}
Adi {Zitrin} and David {Eichler}.
\newblock {Observing Cosmological Processes in Real Time with Repeating Fast Radio Bursts}.
\newblock {\em \apj}, 866(2):101, October 2018.
\newblock \href {http://arxiv.org/abs/1807.03287} {\path{arXiv:1807.03287}}, \href {http://dx.doi.org/10.3847/1538-4357/aad6a2} {\path{[DOI]}}, {\small[\href{https://ui.adsabs.harvard.edu/abs/2018ApJ...866..101Z}{ADS}]}.

\bibitem{LGDWZ2018}
Zheng-Xiang {Li}, He~{Gao}, Xu-Heng {Ding}, Guo-Jian {Wang}, and Bing {Zhang}.
\newblock {Strongly lensed repeating fast radio bursts as precision probes of the universe}.
\newblock {\em Nature Communications}, 9:3833, September 2018.
\newblock \href {http://arxiv.org/abs/1708.06357} {\path{arXiv:1708.06357}}, \href {http://dx.doi.org/10.1038/s41467-018-06303-0} {\path{[DOI]}}, {\small[\href{https://ui.adsabs.harvard.edu/abs/2018NatCo...9.3833L}{ADS}]}.

\bibitem{KL2019}
Pawan {Kumar} and Eric~V. {Linder}.
\newblock {Use of fast radio burst dispersion measures as distance measures}.
\newblock {\em \prd}, 100(8):083533, October 2019.
\newblock \href {http://arxiv.org/abs/1903.08175} {\path{arXiv:1903.08175}}, \href {http://dx.doi.org/10.1103/PhysRevD.100.083533} {\path{[DOI]}}, {\small[\href{https://ui.adsabs.harvard.edu/abs/2019PhRvD.100h3533K}{ADS}]}.

\bibitem{Macquart2020}
J.~P. {Macquart}, J.~X. {Prochaska}, M.~{McQuinn}, K.~W. {Bannister}, S.~{Bhandari}, C.~K. {Day}, A.~T. {Deller}, R.~D. {Ekers}, C.~W. {James}, L.~{Marnoch}, S.~{Os{\l}owski}, C.~{Phillips}, S.~D. {Ryder}, D.~R. {Scott}, R.~M. {Shannon}, and N.~{Tejos}.
\newblock {A census of baryons in the Universe from localized fast radio bursts}.
\newblock {\em \nat}, 581(7809):391--395, May 2020.
\newblock \href {http://arxiv.org/abs/2005.13161} {\path{arXiv:2005.13161}}, \href {http://dx.doi.org/10.1038/s41586-020-2300-2} {\path{[DOI]}}, {\small[\href{https://ui.adsabs.harvard.edu/abs/2020Natur.581..391M}{ADS}]}.

\bibitem{Beniamini2021}
Paz {Beniamini}, Pawan {Kumar}, Xiangcheng {Ma}, and Eliot {Quataert}.
\newblock {Exploring the epoch of hydrogen reionization using FRBs}.
\newblock {\em \mnras}, 502(4):5134--5146, April 2021.
\newblock \href {http://arxiv.org/abs/2011.11643} {\path{arXiv:2011.11643}}, \href {http://dx.doi.org/10.1093/mnras/stab309} {\path{[DOI]}}, {\small[\href{https://ui.adsabs.harvard.edu/abs/2021MNRAS.502.5134B}{ADS}]}.

\bibitem{HR2022}
Steffen {Hagstotz}, Robert {Reischke}, and Robert {Lilow}.
\newblock {A new measurement of the Hubble constant using fast radio bursts}.
\newblock {\em \mnras}, 511(1):662--667, March 2022.
\newblock \href {http://arxiv.org/abs/2104.04538} {\path{arXiv:2104.04538}}, \href {http://dx.doi.org/10.1093/mnras/stac077} {\path{[DOI]}}, {\small[\href{https://ui.adsabs.harvard.edu/abs/2022MNRAS.511..662H}{ADS}]}.

\bibitem{James2022}
C.~W. {James}, E.~M. {Ghosh}, J.~X. {Prochaska}, K.~W. {Bannister}, S.~{Bhandari}, C.~K. {Day}, A.~T. {Deller}, M.~{Glowacki}, A.~C. {Gordon}, K.~E. {Heintz}, L.~{Marnoch}, S.~D. {Ryder}, D.~R. {Scott}, R.~M. {Shannon}, and N.~{Tejos}.
\newblock {A measurement of Hubble's Constant using Fast Radio Bursts}.
\newblock {\em \mnras}, 516(4):4862--4881, November 2022.
\newblock \href {http://arxiv.org/abs/2208.00819} {\path{arXiv:2208.00819}}, \href {http://dx.doi.org/10.1093/mnras/stac2524} {\path{[DOI]}}, {\small[\href{https://ui.adsabs.harvard.edu/abs/2022MNRAS.516.4862J}{ADS}]}.

\bibitem{WZW2022}
Qin {Wu}, Guo-Qiang {Zhang}, and Fa-Yin {Wang}.
\newblock {An 8 per cent determination of the Hubble constant from localized fast radio bursts}.
\newblock {\em \mnras}, 515(1):L1--L5, September 2022.
\newblock \href {http://arxiv.org/abs/2108.00581} {\path{arXiv:2108.00581}}, \href {http://dx.doi.org/10.1093/mnrasl/slac022} {\path{[DOI]}}, {\small[\href{https://ui.adsabs.harvard.edu/abs/2022MNRAS.515L...1W}{ADS}]}.

\bibitem{Zhang2023}
Bing {Zhang}.
\newblock {The physics of fast radio bursts}.
\newblock {\em Reviews of Modern Physics}, 95(3):035005, July 2023.
\newblock \href {http://arxiv.org/abs/2212.03972} {\path{arXiv:2212.03972}}, \href {http://dx.doi.org/10.1103/RevModPhys.95.035005} {\path{[DOI]}}, {\small[\href{https://ui.adsabs.harvard.edu/abs/2023RvMP...95c5005Z}{ADS}]}.

\bibitem{Khrykin2024}
Ilya~S. {Khrykin}, Metin {Ata}, Khee-Gan {Lee}, Sunil {Simha}, Yuxin {Huang}, J.~Xavier {Prochaska}, Nicolas {Tejos}, Keith~W. {Bannister}, Jeff {Cooke}, Cherie~K. {Day}, Adam {Deller}, Marcin {Glowacki}, Alexa~C. {Gordon}, Clancy~W. {James}, Lachlan {Marnoch}, Ryan.~M. {Shannon}, Jielai {Zhang}, and Lucas {Bernales-Cortes}.
\newblock {FLIMFLAM DR1: The First Constraints on the Cosmic Baryon Distribution from 8 FRB sightlines}.
\newblock {\em arXiv e-prints}, page arXiv:2402.00505, February 2024.
\newblock \href {http://arxiv.org/abs/2402.00505} {\path{arXiv:2402.00505}}, \href {http://dx.doi.org/10.48550/arXiv.2402.00505} {\path{[DOI]}}, {\small[\href{https://ui.adsabs.harvard.edu/abs/2024arXiv240200505K}{ADS}]}.

\bibitem{Shaw2024}
Abinash~Kumar {Shaw}, Raghunath {Ghara}, Paz {Beniamini}, Saleem {Zaroubi}, and Pawan {Kumar}.
\newblock {Asking Fast Radio Bursts (FRBs) for More than Reionization History}.
\newblock {\em arXiv e-prints}, page arXiv:2409.03255, September 2024.
\newblock \href {http://arxiv.org/abs/2409.03255} {\path{arXiv:2409.03255}}, \href {http://dx.doi.org/10.48550/arXiv.2409.03255} {\path{[DOI]}}, {\small[\href{https://ui.adsabs.harvard.edu/abs/2024arXiv240903255S}{ADS}]}.

\bibitem{Planck2020}
{Planck Collaboration}.
\newblock {Planck 2018 results. VI. Cosmological parameters}.
\newblock {\em \aap}, 641:A6, September 2020.
\newblock \href {http://arxiv.org/abs/1807.06209} {\path{arXiv:1807.06209}}, \href {http://dx.doi.org/10.1051/0004-6361/201833910} {\path{[DOI]}}, {\small[\href{https://ui.adsabs.harvard.edu/abs/2020A&A...641A...6P}{ADS}]}.

\bibitem{Riess2019}
Adam~G. {Riess}, Stefano {Casertano}, Wenlong {Yuan}, Lucas~M. {Macri}, and Dan {Scolnic}.
\newblock {Large Magellanic Cloud Cepheid Standards Provide a 1\% Foundation for the Determination of the Hubble Constant and Stronger Evidence for Physics beyond {\ensuremath{\Lambda}}CDM}.
\newblock {\em \apj}, 876(1):85, May 2019.
\newblock \href {http://arxiv.org/abs/1903.07603} {\path{arXiv:1903.07603}}, \href {http://dx.doi.org/10.3847/1538-4357/ab1422} {\path{[DOI]}}, {\small[\href{https://ui.adsabs.harvard.edu/abs/2019ApJ...876...85R}{ADS}]}.

\bibitem{Verde2019}
Licia {Verde}, Tommaso {Treu}, and Adam~G. {Riess}.
\newblock {Tensions between the early and late Universe}.
\newblock {\em Nature Astronomy}, 3:891--895, September 2019.
\newblock \href {http://arxiv.org/abs/1907.10625} {\path{arXiv:1907.10625}}, \href {http://dx.doi.org/10.1038/s41550-019-0902-0} {\path{[DOI]}}, {\small[\href{https://ui.adsabs.harvard.edu/abs/2019NatAs...3..891V}{ADS}]}.

\bibitem{Walters2018}
Anthony {Walters}, Amanda {Weltman}, B.~M. {Gaensler}, Yin-Zhe {Ma}, and Amadeus {Witzemann}.
\newblock {Future Cosmological Constraints From Fast Radio Bursts}.
\newblock {\em \apj}, 856(1):65, March 2018.
\newblock \href {http://arxiv.org/abs/1711.11277} {\path{arXiv:1711.11277}}, \href {http://dx.doi.org/10.3847/1538-4357/aaaf6b} {\path{[DOI]}}, {\small[\href{https://ui.adsabs.harvard.edu/abs/2018ApJ...856...65W}{ADS}]}.

\bibitem{Zhao2020}
Ze-Wei {Zhao}, Zheng-Xiang {Li}, Jing-Zhao {Qi}, He~{Gao}, Jing-Fei {Zhang}, and Xin {Zhang}.
\newblock {Cosmological Parameter Estimation for Dynamical Dark Energy Models with Future Fast Radio Burst Observations}.
\newblock {\em \apj}, 903(2):83, November 2020.
\newblock \href {http://arxiv.org/abs/2006.01450} {\path{arXiv:2006.01450}}, \href {http://dx.doi.org/10.3847/1538-4357/abb8ce} {\path{[DOI]}}, {\small[\href{https://ui.adsabs.harvard.edu/abs/2020ApJ...903...83Z}{ADS}]}.

\bibitem{Hagstotz2022}
Steffen {Hagstotz}, Robert {Reischke}, and Robert {Lilow}.
\newblock {A new measurement of the Hubble constant using fast radio bursts}.
\newblock {\em \mnras}, 511(1):662--667, March 2022.
\newblock \href {http://arxiv.org/abs/2104.04538} {\path{arXiv:2104.04538}}, \href {http://dx.doi.org/10.1093/mnras/stac077} {\path{[DOI]}}, {\small[\href{https://ui.adsabs.harvard.edu/abs/2022MNRAS.511..662H}{ADS}]}.

\bibitem{Wu2022}
Qin {Wu}, Guo-Qiang {Zhang}, and Fa-Yin {Wang}.
\newblock {An 8 per cent determination of the Hubble constant from localized fast radio bursts}.
\newblock {\em \mnras}, 515(1):L1--L5, September 2022.
\newblock \href {http://arxiv.org/abs/2108.00581} {\path{arXiv:2108.00581}}, \href {http://dx.doi.org/10.1093/mnrasl/slac022} {\path{[DOI]}}, {\small[\href{https://ui.adsabs.harvard.edu/abs/2022MNRAS.515L...1W}{ADS}]}.

\bibitem{Liu2023}
Yang {Liu}, Hongwei {Yu}, and Puxun {Wu}.
\newblock {Cosmological-model-independent Determination of Hubble Constant from Fast Radio Bursts and Hubble Parameter Measurements}.
\newblock {\em \apjl}, 946(2):L49, April 2023.
\newblock \href {http://arxiv.org/abs/2210.05202} {\path{arXiv:2210.05202}}, \href {http://dx.doi.org/10.3847/2041-8213/acc650} {\path{[DOI]}}, {\small[\href{https://ui.adsabs.harvard.edu/abs/2023ApJ...946L..49L}{ADS}]}.

\bibitem{Fortunato2023}
J{\'e}ferson A.~S. {Fortunato}, Wiliam~S. {Hip{\'o}lito-Ricaldi}, and Marcelo~V. {dos Santos}.
\newblock {Cosmography from well-localized fast radio bursts}.
\newblock {\em \mnras}, 526(2):1773--1782, December 2023.
\newblock \href {http://arxiv.org/abs/2307.04711} {\path{arXiv:2307.04711}}, \href {http://dx.doi.org/10.1093/mnras/stad2856} {\path{[DOI]}}, {\small[\href{https://ui.adsabs.harvard.edu/abs/2023MNRAS.526.1773F}{ADS}]}.

\bibitem{AB2025}
Sandeep~Kumar {Acharya} and Paz {Beniamini}.
\newblock {Redshift dependence of FRB host dispersion measures across cosmic epochs}.
\newblock {\em \jcap}, 2025(1):036, January 2025.
\newblock \href {http://arxiv.org/abs/2408.03163} {\path{arXiv:2408.03163}}, \href {http://dx.doi.org/10.1088/1475-7516/2025/01/036} {\path{[DOI]}}, {\small[\href{https://ui.adsabs.harvard.edu/abs/2025JCAP...01..036A}{ADS}]}.

\bibitem{Mo2023}
Jian-Feng {Mo}, Weishan {Zhu}, Yang {Wang}, Lin {Tang}, and Long-Long {Feng}.
\newblock {The dispersion measure of Fast Radio Bursts host galaxies: estimation from cosmological simulations}.
\newblock {\em \mnras}, 518(1):539--561, January 2023.
\newblock \href {http://arxiv.org/abs/2210.14052} {\path{arXiv:2210.14052}}, \href {http://dx.doi.org/10.1093/mnras/stac3104} {\path{[DOI]}}, {\small[\href{https://ui.adsabs.harvard.edu/abs/2023MNRAS.518..539M}{ADS}]}.

\bibitem{IllustrisTNG}
Annalisa {Pillepich}, Volker {Springel}, Dylan {Nelson}, Shy {Genel}, Jill {Naiman}, R{\"u}diger {Pakmor}, Lars {Hernquist}, Paul {Torrey}, Mark {Vogelsberger}, Rainer {Weinberger}, and Federico {Marinacci}.
\newblock {Simulating galaxy formation with the IllustrisTNG model}.
\newblock {\em \mnras}, 473(3):4077--4106, January 2018.
\newblock \href {http://arxiv.org/abs/1703.02970} {\path{arXiv:1703.02970}}, \href {http://dx.doi.org/10.1093/mnras/stx2656} {\path{[DOI]}}, {\small[\href{https://ui.adsabs.harvard.edu/abs/2018MNRAS.473.4077P}{ADS}]}.

\bibitem{CoDa2020}
Pierre {Ocvirk}, Dominique {Aubert}, Jenny~G. {Sorce}, Paul~R. {Shapiro}, Nicolas {Deparis}, Taha {Dawoodbhoy}, Joseph {Lewis}, Romain {Teyssier}, Gustavo {Yepes}, Stefan {Gottl{\"o}ber}, Kyungjin {Ahn}, Ilian~T. {Iliev}, and Yehuda {Hoffman}.
\newblock {Cosmic Dawn II (CoDa II): a new radiation-hydrodynamics simulation of the self-consistent coupling of galaxy formation and reionization}.
\newblock {\em \mnras}, 496(4):4087--4107, August 2020.
\newblock \href {http://arxiv.org/abs/1811.11192} {\path{arXiv:1811.11192}}, \href {http://dx.doi.org/10.1093/mnras/staa1266} {\path{[DOI]}}, {\small[\href{https://ui.adsabs.harvard.edu/abs/2020MNRAS.496.4087O}{ADS}]}.

\bibitem{DMGNK2021}
Sanskriti {Das}, Smita {Mathur}, Anjali {Gupta}, Fabrizio {Nicastro}, and Yair {Krongold}.
\newblock {Empirical estimates of the Galactic halo contribution to the dispersion measures of extragalactic fast radio bursts using X-ray absorption}.
\newblock {\em \mnras}, 500(1):655--662, January 2021.
\newblock \href {http://arxiv.org/abs/2007.11542} {\path{arXiv:2007.11542}}, \href {http://dx.doi.org/10.1093/mnras/staa3299} {\path{[DOI]}}, {\small[\href{https://ui.adsabs.harvard.edu/abs/2021MNRAS.500..655D}{ADS}]}.

\bibitem{CL2002}
J.~M. {Cordes} and T.~J.~W. {Lazio}.
\newblock {NE2001.I. A New Model for the Galactic Distribution of Free Electrons and its Fluctuations}.
\newblock {\em arXiv e-prints}, page 0207156, July 2002.
\newblock \href {http://arxiv.org/abs/0207156} {\path{arXiv:0207156}}, \href {http://dx.doi.org/10.48550/arXiv.astro-ph/0207156} {\path{[DOI]}}, {\small[\href{https://ui.adsabs.harvard.edu/abs/2002astro.ph..7156C}{ADS}]}.

\bibitem{Yao2017}
J.~M. {Yao}, R.~N. {Manchester}, and N.~{Wang}.
\newblock {A New Electron-density Model for Estimation of Pulsar and FRB Distances}.
\newblock {\em \apj}, 835(1):29, January 2017.
\newblock \href {http://arxiv.org/abs/1610.09448} {\path{arXiv:1610.09448}}, \href {http://dx.doi.org/10.3847/1538-4357/835/1/29} {\path{[DOI]}}, {\small[\href{https://ui.adsabs.harvard.edu/abs/2017ApJ...835...29Y}{ADS}]}.

\bibitem{Hilmarsson2021}
G.~H. {Hilmarsson}, D.~{Michilli}, L.~G. {Spitler}, R.~S. {Wharton}, P.~{Demorest}, G.~{Desvignes}, K.~{Gourdji}, S.~{Hackstein}, J.~W.~T. {Hessels}, K.~{Nimmo}, A.~D. {Seymour}, M.~{Kramer}, and R.~{Mckinven}.
\newblock {Rotation Measure Evolution of the Repeating Fast Radio Burst Source FRB 121102}.
\newblock {\em \apjl}, 908(1):L10, February 2021.
\newblock \href {http://arxiv.org/abs/2009.12135} {\path{arXiv:2009.12135}}, \href {http://dx.doi.org/10.3847/2041-8213/abdec0} {\path{[DOI]}}, {\small[\href{https://ui.adsabs.harvard.edu/abs/2021ApJ...908L..10H}{ADS}]}.

\bibitem{AnnaThomas2023}
Reshma {Anna-Thomas et al}.
\newblock {Magnetic field reversal in the turbulent environment around a repeating fast radio burst}.
\newblock {\em Science}, 380(6645):599--603, May 2023.
\newblock \href {http://arxiv.org/abs/2202.11112} {\path{arXiv:2202.11112}}, \href {http://dx.doi.org/10.1126/science.abo6526} {\path{[DOI]}}, {\small[\href{https://ui.adsabs.harvard.edu/abs/2023Sci...380..599A}{ADS}]}.

\bibitem{Chatterjee2017}
S.~{Chatterjee}, C.~J. {Law}, R.~S. {Wharton}, S.~{Burke-Spolaor}, J.~W.~T. {Hessels}, G.~C. {Bower}, J.~M. {Cordes}, S.~P. {Tendulkar}, C.~G. {Bassa}, P.~{Demorest}, B.~J. {Butler}, A.~{Seymour}, P.~{Scholz}, M.~W. {Abruzzo}, S.~{Bogdanov}, V.~M. {Kaspi}, A.~{Keimpema}, T.~J.~W. {Lazio}, B.~{Marcote}, M.~A. {McLaughlin}, Z.~{Paragi}, S.~M. {Ransom}, M.~{Rupen}, L.~G. {Spitler}, and H.~J. {van Langevelde}.
\newblock {A direct localization of a fast radio burst and its host}.
\newblock {\em \nat}, 541(7635):58--61, January 2017.
\newblock \href {http://arxiv.org/abs/1701.01098} {\path{arXiv:1701.01098}}, \href {http://dx.doi.org/10.1038/nature20797} {\path{[DOI]}}, {\small[\href{https://ui.adsabs.harvard.edu/abs/2017Natur.541...58C}{ADS}]}.

\bibitem{Niu2022}
C.~H. {Niu}, K.~{Aggarwal}, D.~{Li}, X.~{Zhang}, S.~{Chatterjee}, C.~W. {Tsai}, W.~{Yu}, C.~J. {Law}, S.~{Burke-Spolaor}, J.~M. {Cordes}, Y.~K. {Zhang}, S.~K. {Ocker}, J.~M. {Yao}, P.~{Wang}, Y.~{Feng}, Y.~{Niino}, C.~{Bochenek}, M.~{Cruces}, L.~{Connor}, J.~A. {Jiang}, S.~{Dai}, R.~{Luo}, G.~D. {Li}, C.~C. {Miao}, J.~R. {Niu}, R.~{Anna-Thomas}, J.~{Sydnor}, D.~{Stern}, W.~Y. {Wang}, M.~{Yuan}, Y.~L. {Yue}, D.~J. {Zhou}, Z.~{Yan}, W.~W. {Zhu}, and B.~{Zhang}.
\newblock {A repeating fast radio burst associated with a persistent radio source}.
\newblock {\em \nat}, 606(7916):873--877, June 2022.
\newblock \href {http://arxiv.org/abs/2110.07418} {\path{arXiv:2110.07418}}, \href {http://dx.doi.org/10.1038/s41586-022-04755-5} {\path{[DOI]}}, {\small[\href{https://ui.adsabs.harvard.edu/abs/2022Natur.606..873N}{ADS}]}.

\bibitem{Mcquinn2016}
Matthew {McQuinn}.
\newblock {The Evolution of the Intergalactic Medium}.
\newblock {\em \araa}, 54:313--362, September 2016.
\newblock \href {http://arxiv.org/abs/1512.00086} {\path{arXiv:1512.00086}}, \href {http://dx.doi.org/10.1146/annurev-astro-082214-122355} {\path{[DOI]}}, {\small[\href{https://ui.adsabs.harvard.edu/abs/2016ARA&A..54..313M}{ADS}]}.

\bibitem{Ziegler2024}
Joshua~J. {Ziegler}, Paul~R. {Shapiro}, Taha {Dawoodbhoy}, Paz {Beniamini}, Pawan {Kumar}, Katherine {Freese}, Pierre {Ocvirk}, Dominique {Aubert}, Joseph S.~W. {Lewis}, Romain {Teyssier}, Hyunbae {Park}, Kyungjin {Ahn}, Jenny~G. {Sorce}, Ilian~T. {Iliev}, Gustavo {Yepes}, and Stefan {Gottlober}.
\newblock {Dispersion Measures of Fast Radio Bursts through the Epoch of Reionization}.
\newblock {\em arXiv e-prints}, page arXiv:2411.02699, November 2024.
\newblock \href {http://arxiv.org/abs/2411.02699} {\path{arXiv:2411.02699}}, \href {http://dx.doi.org/10.48550/arXiv.2411.02699} {\path{[DOI]}}, {\small[\href{https://ui.adsabs.harvard.edu/abs/2024arXiv241102699Z}{ADS}]}.

\bibitem{Zhang2021}
Z.~J. {Zhang}, K.~{Yan}, C.~M. {Li}, G.~Q. {Zhang}, and F.~Y. {Wang}.
\newblock {Intergalactic Medium Dispersion Measures of Fast Radio Bursts Estimated from IllustrisTNG Simulation and Their Cosmological Applications}.
\newblock {\em \apj}, 906(1):49, January 2021.
\newblock \href {http://arxiv.org/abs/2011.14494} {\path{arXiv:2011.14494}}, \href {http://dx.doi.org/10.3847/1538-4357/abceb9} {\path{[DOI]}}, {\small[\href{https://ui.adsabs.harvard.edu/abs/2021ApJ...906...49Z}{ADS}]}.

\bibitem{Zhang2020}
G.~Q. {Zhang}, Hai {Yu}, J.~H. {He}, and F.~Y. {Wang}.
\newblock {Dispersion Measures of Fast Radio Burst Host Galaxies Derived from IllustrisTNG Simulation}.
\newblock {\em \apj}, 900(2):170, September 2020.
\newblock \href {http://arxiv.org/abs/2007.13935} {\path{arXiv:2007.13935}}, \href {http://dx.doi.org/10.3847/1538-4357/abaa4a} {\path{[DOI]}}, {\small[\href{https://ui.adsabs.harvard.edu/abs/2020ApJ...900..170Z}{ADS}]}.

\bibitem{Nelson2019}
Dylan {Nelson}, Volker {Springel}, Annalisa {Pillepich}, Vicente {Rodriguez-Gomez}, Paul {Torrey}, Shy {Genel}, Mark {Vogelsberger}, Ruediger {Pakmor}, Federico {Marinacci}, Rainer {Weinberger}, Luke {Kelley}, Mark {Lovell}, Benedikt {Diemer}, and Lars {Hernquist}.
\newblock {The IllustrisTNG simulations: public data release}.
\newblock {\em Computational Astrophysics and Cosmology}, 6(1):2, May 2019.
\newblock \href {http://arxiv.org/abs/1812.05609} {\path{arXiv:1812.05609}}, \href {http://dx.doi.org/10.1186/s40668-019-0028-x} {\path{[DOI]}}, {\small[\href{https://ui.adsabs.harvard.edu/abs/2019ComAC...6....2N}{ADS}]}.

\bibitem{ME2018}
J.~P. {Macquart} and RD~{Ekers}.
\newblock {FRB event rate counts - II. Fluence, redshift, and dispersion measure distributions}.
\newblock {\em \mnras}, 480(3):4211--4230, November 2018.
\newblock \href {http://arxiv.org/abs/1808.00908} {\path{arXiv:1808.00908}}, \href {http://dx.doi.org/10.1093/mnras/sty2083} {\path{[DOI]}}, {\small[\href{https://ui.adsabs.harvard.edu/abs/2018MNRAS.480.4211M}{ADS}]}.

\bibitem{DSA}
Gregg {Hallinan}, Vikram {Ravi}, Katie {Bouman}, Fabian {Walter}, Francois {Kapp}, Katie {Jameson}, and {DSA-2000 collaboration}.
\newblock {The DSA-2000 Radio Camera}.
\newblock In {\em American Astronomical Society Meeting Abstracts}, volume 243 of {\em American Astronomical Society Meeting Abstracts}, page 237.05, February 2024.
\newblock {\small[\href{https://ui.adsabs.harvard.edu/abs/2024AAS...24323705H}{ADS}]}.

\bibitem{CHORD}
Keith {Vanderlinde}, Adrian {Liu}, Bryan {Gaensler}, Dick {Bond}, Gary {Hinshaw}, Cherry {Ng}, Cynthia {Chiang}, Ingrid {Stairs}, Jo-Anne {Brown}, Jonathan {Sievers}, Juan {Mena}, Kendrick {Smith}, Kevin {Bandura}, Kiyoshi {Masui}, Kristine {Spekkens}, Leo {Belostotski}, Matt {Dobbs}, Neil {Turok}, Patrick {Boyle}, Michael {Rupen}, Tom {Landecker}, Ue-Li {Pen}, and Victoria {Kaspi}.
\newblock {The Canadian Hydrogen Observatory and Radio-transient Detector (CHORD)}.
\newblock In {\em Canadian Long Range Plan for Astronomy and Astrophysics White Papers}, volume 2020, page~28, October 2019.
\newblock \href {http://arxiv.org/abs/1911.01777} {\path{arXiv:1911.01777}}, \href {http://dx.doi.org/10.5281/zenodo.3765414} {\path{[DOI]}}, {\small[\href{https://ui.adsabs.harvard.edu/abs/2019clrp.2020...28V}{ADS}]}.

\bibitem{Bhardwajselection}
Mohit {Bhardwaj}, Jimin {Lee}, and Kevin {Ji}.
\newblock {Selection bias obfuscates the discovery of fast radio burst sources}.
\newblock {\em \nat}, 634(8036):1065--1069, October 2024.
\newblock \href {http://arxiv.org/abs/2408.01876} {\path{arXiv:2408.01876}}, \href {http://dx.doi.org/10.1038/s41586-024-08065-w} {\path{[DOI]}}, {\small[\href{https://ui.adsabs.harvard.edu/abs/2024Natur.634.1065B}{ADS}]}.

\bibitem{Glowacki2025}
M.~{Glowacki}, A.~{Bera}, C.~W. {James}, J.~{Paterson}, A.~T. {Deller}, A~C. {Gordon}, L.~{Marnoch}, A.~R. {Muller}, J.~X. {Prochaska}, S.~D. {Ryder}, R.~M. {Shannon}, N.~{Tejos}, and A.~G. {Mannings}.
\newblock {An investigation into correlations between FRB and host galaxy properties}.
\newblock {\em arXiv e-prints}, page arXiv:2506.23403, June 2025.
\newblock \href {http://arxiv.org/abs/2506.23403} {\path{arXiv:2506.23403}}, \href {http://dx.doi.org/10.48550/arXiv.2506.23403} {\path{[DOI]}}, {\small[\href{https://ui.adsabs.harvard.edu/abs/2025arXiv250623403G}{ADS}]}.

\bibitem{HPCFL2007}
F.~{Hammer}, M.~{Puech}, L.~{Chemin}, H.~{Flores}, and M.~D. {Lehnert}.
\newblock {The Milky Way, an Exceptionally Quiet Galaxy: Implications for the Formation of Spiral Galaxies}.
\newblock {\em \apj}, 662(1):322--334, June 2007.
\newblock \href {http://arxiv.org/abs/astro-ph/0702585} {\path{arXiv:astro-ph/0702585}}, \href {http://dx.doi.org/10.1086/516727} {\path{[DOI]}}, {\small[\href{https://ui.adsabs.harvard.edu/abs/2007ApJ...662..322H}{ADS}]}.

\bibitem{Sharma2024}
Kritti {Sharma et al}.
\newblock {Preferential Occurrence of Fast Radio Bursts in Massive Star-Forming Galaxies}.
\newblock {\em arXiv e-prints}, page arXiv:2409.16964, September 2024.
\newblock \href {http://arxiv.org/abs/2409.16964} {\path{arXiv:2409.16964}}, \href {http://dx.doi.org/10.48550/arXiv.2409.16964} {\path{[DOI]}}, {\small[\href{https://ui.adsabs.harvard.edu/abs/2024arXiv240916964S}{ADS}]}.

\bibitem{Ma2018}
Xiangcheng {Ma}, Philip~F. {Hopkins}, Shea {Garrison-Kimmel}, Claude-Andr{\'e} {Faucher-Gigu{\`e}re}, Eliot {Quataert}, Michael {Boylan-Kolchin}, Christopher~C. {Hayward}, Robert {Feldmann}, and Du{\v{s}}an {Kere{\v{s}}}.
\newblock {Simulating galaxies in the reionization era with FIRE-2: galaxy scaling relations, stellar mass functions, and luminosity functions}.
\newblock {\em \mnras}, 478(2):1694--1715, August 2018.
\newblock \href {http://arxiv.org/abs/1706.06605} {\path{arXiv:1706.06605}}, \href {http://dx.doi.org/10.1093/mnras/sty1024} {\path{[DOI]}}, {\small[\href{https://ui.adsabs.harvard.edu/abs/2018MNRAS.478.1694M}{ADS}]}.

\bibitem{Jahns2023}
J.~N. {Jahns}, L.~G. {Spitler}, K.~{Nimmo}, D.~M. {Hewitt}, M.~P. {Snelders}, A.~{Seymour}, J.~W.~T. {Hessels}, K.~{Gourdji}, D.~{Michilli}, and G.~H. {Hilmarsson}.
\newblock {The FRB 20121102A November rain in 2018 observed with the Arecibo Telescope}.
\newblock {\em \mnras}, 519(1):666--687, February 2023.
\newblock \href {http://arxiv.org/abs/2202.05705} {\path{arXiv:2202.05705}}, \href {http://dx.doi.org/10.1093/mnras/stac3446} {\path{[DOI]}}, {\small[\href{https://ui.adsabs.harvard.edu/abs/2023MNRAS.519..666J}{ADS}]}.

\bibitem{RABG2025}
Sk.~Minhajur {Rahaman}, Sandeep~Kumar {Acharya}, Paz {Beniamini}, and Jonathan {Granot}.
\newblock {Persistent Radio Sources Associated with Fast Radio Bursts: Implications from Magnetar Progenitors}.
\newblock {\em \apj}, 988(2):276, August 2025.
\newblock \href {http://arxiv.org/abs/2504.01125} {\path{arXiv:2504.01125}}, \href {http://dx.doi.org/10.3847/1538-4357/ade70c} {\path{[DOI]}}, {\small[\href{https://ui.adsabs.harvard.edu/abs/2025ApJ...988..276R}{ADS}]}.

\bibitem{Bhardwaj2021}
M.~{Bhardwaj}, A.~Yu. {Kirichenko}, D.~{Michilli}, Y.~D. {Mayya}, V.~M. {Kaspi}, B.~M. {Gaensler}, M.~{Rahman}, S.~P. {Tendulkar}, E.~{Fonseca}, Alexander {Josephy}, C.~{Leung}, Marcus {Merryfield}, Emily {Petroff}, Z.~{Pleunis}, Pranav {Sanghavi}, P.~{Scholz}, K.~{Shin}, Kendrick~M. {Smith}, and I.~H. {Stairs}.
\newblock {A Local Universe Host for the Repeating Fast Radio Burst FRB 20181030A}.
\newblock {\em \apjl}, 919(2):L24, October 2021.
\newblock \href {http://arxiv.org/abs/2108.12122} {\path{arXiv:2108.12122}}, \href {http://dx.doi.org/10.3847/2041-8213/ac223b} {\path{[DOI]}}, {\small[\href{https://ui.adsabs.harvard.edu/abs/2021ApJ...919L..24B}{ADS}]}.

\bibitem{Kirsten2022}
F.~{Kirsten et al}.
\newblock {A repeating fast radio burst source in a globular cluster}.
\newblock {\em \nat}, 602(7898):585--589, February 2022.
\newblock \href {http://arxiv.org/abs/2105.11445} {\path{arXiv:2105.11445}}, \href {http://dx.doi.org/10.1038/s41586-021-04354-w} {\path{[DOI]}}, {\small[\href{https://ui.adsabs.harvard.edu/abs/2022Natur.602..585K}{ADS}]}.

\bibitem{Ravi2019_1}
Vikram {Ravi}.
\newblock {The observed properties of fast radio bursts}.
\newblock {\em \mnras}, 482(2):1966--1978, January 2019.
\newblock \href {http://arxiv.org/abs/1710.08026} {\path{arXiv:1710.08026}}, \href {http://dx.doi.org/10.1093/mnras/sty1551} {\path{[DOI]}}, {\small[\href{https://ui.adsabs.harvard.edu/abs/2019MNRAS.482.1966R}{ADS}]}.

\bibitem{GFFBDDL2022}
V.~{Gupta}, C.~{Flynn}, W.~{Farah}, M.~{Bailes}, A.~T. {Deller}, C.~K. {Day}, and M.~E. {Lower}.
\newblock {The ultranarrow FRB20191107B, and the origins of FRB scattering}.
\newblock {\em \mnras}, 514(4):5866--5878, August 2022.
\newblock \href {http://arxiv.org/abs/2209.00311} {\path{arXiv:2209.00311}}, \href {http://dx.doi.org/10.1093/mnras/stac1720} {\path{[DOI]}}, {\small[\href{https://ui.adsabs.harvard.edu/abs/2022MNRAS.514.5866G}{ADS}]}.

\bibitem{BK2020}
Paz {Beniamini} and Pawan {Kumar}.
\newblock {What does FRB light-curve variability tell us about the emission mechanism?}
\newblock {\em \mnras}, 498(1):651--664, August 2020.
\newblock \href {http://arxiv.org/abs/2007.07265} {\path{arXiv:2007.07265}}, \href {http://dx.doi.org/10.1093/mnras/staa2489} {\path{[DOI]}}, {\small[\href{https://ui.adsabs.harvard.edu/abs/2020MNRAS.498..651B}{ADS}]}.

\bibitem{Chime_catalog}
{CHIME/FRB Collaboration} and {Amiri et al}.
\newblock {The First CHIME/FRB Fast Radio Burst Catalog}.
\newblock {\em \apjs}, 257(2):59, December 2021.
\newblock \href {http://arxiv.org/abs/2106.04352} {\path{arXiv:2106.04352}}, \href {http://dx.doi.org/10.3847/1538-4365/ac33ab} {\path{[DOI]}}, {\small[\href{https://ui.adsabs.harvard.edu/abs/2021ApJS..257...59C}{ADS}]}.

\bibitem{Cordes2016}
J.~M. {Cordes}, R.~S. {Wharton}, L.~G. {Spitler}, S.~{Chatterjee}, and I.~{Wasserman}.
\newblock {Radio Wave Propagation and the Provenance of Fast Radio Bursts}.
\newblock {\em arXiv e-prints}, page arXiv:1605.05890, May 2016.
\newblock \href {http://arxiv.org/abs/1605.05890} {\path{arXiv:1605.05890}}, \href {http://dx.doi.org/10.48550/arXiv.1605.05890} {\path{[DOI]}}, {\small[\href{https://ui.adsabs.harvard.edu/abs/2016arXiv160505890C}{ADS}]}.

\bibitem{Marcote2020}
B.~{Marcote et al}.
\newblock {A repeating fast radio burst source localized to a nearby spiral galaxy}.
\newblock {\em \nat}, 577(7789):190--194, January 2020.
\newblock \href {http://arxiv.org/abs/2001.02222} {\path{arXiv:2001.02222}}, \href {http://dx.doi.org/10.1038/s41586-019-1866-z} {\path{[DOI]}}, {\small[\href{https://ui.adsabs.harvard.edu/abs/2020Natur.577..190M}{ADS}]}.

\bibitem{Rajwade2022}
K.~M. {Rajwade et al}.
\newblock {First discoveries and localizations of Fast Radio Bursts with MeerTRAP: real-time, commensal MeerKAT survey}.
\newblock {\em \mnras}, 514(2):1961--1974, August 2022.
\newblock \href {http://arxiv.org/abs/2205.14600} {\path{arXiv:2205.14600}}, \href {http://dx.doi.org/10.1093/mnras/stac1450} {\path{[DOI]}}, {\small[\href{https://ui.adsabs.harvard.edu/abs/2022MNRAS.514.1961R}{ADS}]}.

\bibitem{Driessen2024}
L.~N. {Driessen et al}.
\newblock {FRB 20210405I: a nearby Fast Radio Burst localized to sub-arcsecond precision with MeerKAT}.
\newblock {\em \mnras}, 527(2):3659--3673, January 2024.
\newblock \href {http://arxiv.org/abs/2302.09787} {\path{arXiv:2302.09787}}, \href {http://dx.doi.org/10.1093/mnras/stad3329} {\path{[DOI]}}, {\small[\href{https://ui.adsabs.harvard.edu/abs/2024MNRAS.527.3659D}{ADS}]}.

\bibitem{Connor2023}
Liam {Connor et al}.
\newblock {Deep Synoptic Array Science: Two Fast Radio Burst Sources in Massive Galaxy Clusters}.
\newblock {\em \apjl}, 949(2):L26, June 2023.
\newblock \href {http://arxiv.org/abs/2302.14788} {\path{arXiv:2302.14788}}, \href {http://dx.doi.org/10.3847/2041-8213/acd3ea} {\path{[DOI]}}, {\small[\href{https://ui.adsabs.harvard.edu/abs/2023ApJ...949L..26C}{ADS}]}.

\bibitem{Day2020}
Cherie~K. {Day}, Adam~T. {Deller}, Ryan~M. {Shannon}, Hao {Qiu}, Keith~W. {Bannister}, Shivani {Bhandari}, Ron {Ekers}, Chris {Flynn}, Clancy~W. {James}, Jean-Pierre {Macquart}, Elizabeth~K. {Mahony}, Chris~J. {Phillips}, and J.~{Xavier Prochaska}.
\newblock {High time resolution and polarization properties of ASKAP-localized fast radio bursts}.
\newblock {\em \mnras}, 497(3):3335--3350, September 2020.
\newblock \href {http://arxiv.org/abs/2005.13162} {\path{arXiv:2005.13162}}, \href {http://dx.doi.org/10.1093/mnras/staa2138} {\path{[DOI]}}, {\small[\href{https://ui.adsabs.harvard.edu/abs/2020MNRAS.497.3335D}{ADS}]}.

\bibitem{Eftekhari2025}
T.~{Eftekhari et al}.
\newblock {The Massive and Quiescent Elliptical Host Galaxy of the Repeating Fast Radio Burst FRB 20240209A}.
\newblock {\em \apjl}, 979(2):L22, February 2025.
\newblock \href {http://arxiv.org/abs/2410.23336} {\path{arXiv:2410.23336}}, \href {http://dx.doi.org/10.3847/2041-8213/ad9de2} {\path{[DOI]}}, {\small[\href{https://ui.adsabs.harvard.edu/abs/2025ApJ...979L..22E}{ADS}]}.

\bibitem{Shah2025}
Vishwangi {Shah et al}.
\newblock {A Repeating Fast Radio Burst Source in the Outskirts of a Quiescent Galaxy}.
\newblock {\em \apjl}, 979(2):L21, February 2025.
\newblock \href {http://arxiv.org/abs/2410.23374} {\path{arXiv:2410.23374}}, \href {http://dx.doi.org/10.3847/2041-8213/ad9ddc} {\path{[DOI]}}, {\small[\href{https://ui.adsabs.harvard.edu/abs/2025ApJ...979L..21S}{ADS}]}.

\bibitem{Caleb2023}
M.~{Caleb et al}.
\newblock {A subarcsec localized fast radio burst with a significant host galaxy dispersion measure contribution}.
\newblock {\em \mnras}, 524(2):2064--2077, September 2023.
\newblock \href {http://arxiv.org/abs/2302.09754} {\path{arXiv:2302.09754}}, \href {http://dx.doi.org/10.1093/mnras/stad1839} {\path{[DOI]}}, {\small[\href{https://ui.adsabs.harvard.edu/abs/2023MNRAS.524.2064C}{ADS}]}.

\bibitem{Cassanelli2024}
Tomas {Cassanelli et al}.
\newblock {A fast radio burst localized at detection to an edge-on galaxy using very-long-baseline interferometry}.
\newblock {\em Nature Astronomy}, 8:1429--1442, November 2024.
\newblock \href {http://arxiv.org/abs/2307.09502} {\path{arXiv:2307.09502}}, \href {http://dx.doi.org/10.1038/s41550-024-02357-x} {\path{[DOI]}}, {\small[\href{https://ui.adsabs.harvard.edu/abs/2024NatAs...8.1429C}{ADS}]}.

\bibitem{Ocker2021}
Stella~Koch {Ocker}, James~M. {Cordes}, and Shami {Chatterjee}.
\newblock {Constraining Galaxy Halos from the Dispersion and Scattering of Fast Radio Bursts and Pulsars}.
\newblock {\em \apj}, 911(2):102, April 2021.
\newblock \href {http://arxiv.org/abs/2101.04784} {\path{arXiv:2101.04784}}, \href {http://dx.doi.org/10.3847/1538-4357/abeb6e} {\path{[DOI]}}, {\small[\href{https://ui.adsabs.harvard.edu/abs/2021ApJ...911..102O}{ADS}]}.

\bibitem{Bhandari2023}
Shivani {Bhandari}, Alexa~C. {Gordon}, Danica~R. {Scott}, Lachlan {Marnoch}, Navin {Sridhar}, Pravir {Kumar}, Clancy~W. {James}, Hao {Qiu}, Keith~W. {Bannister}, Adam {T. Deller}, Tarraneh {Eftekhari}, Wen-fai {Fong}, Marcin {Glowacki}, J.~Xavier {Prochaska}, Stuart~D. {Ryder}, Ryan~M. {Shannon}, and Sunil {Simha}.
\newblock {A Nonrepeating Fast Radio Burst in a Dwarf Host Galaxy}.
\newblock {\em \apj}, 948(1):67, May 2023.
\newblock \href {http://arxiv.org/abs/2211.16790} {\path{arXiv:2211.16790}}, \href {http://dx.doi.org/10.3847/1538-4357/acc178} {\path{[DOI]}}, {\small[\href{https://ui.adsabs.harvard.edu/abs/2023ApJ...948...67B}{ADS}]}.

\bibitem{Bhandari2020}
Shivani {Bhandari}, Keith~W. {Bannister}, Emil {Lenc}, Hyerin {Cho}, Ron {Ekers}, Cherie~K. {Day}, Adam~T. {Deller}, Chris {Flynn}, Clancy~W. {James}, Jean-Pierre {Macquart}, Elizabeth~K. {Mahony}, Lachlan {Marnoch}, Vanessa~A. {Moss}, Chris {Phillips}, J.~Xavier {Prochaska}, Hao {Qiu}, Stuart~D. {Ryder}, Ryan~M. {Shannon}, Nicolas {Tejos}, and O.~Ivy {Wong}.
\newblock {Limits on Precursor and Afterglow Radio Emission from a Fast Radio Burst in a Star-forming Galaxy}.
\newblock {\em \apjl}, 901(2):L20, October 2020.
\newblock \href {http://arxiv.org/abs/2008.12488} {\path{arXiv:2008.12488}}, \href {http://dx.doi.org/10.3847/2041-8213/abb462} {\path{[DOI]}}, {\small[\href{https://ui.adsabs.harvard.edu/abs/2020ApJ...901L..20B}{ADS}]}.

\bibitem{Qiu2020}
Hao {Qiu}, Ryan~M. {Shannon}, Wael {Farah}, Jean-Pierre {Macquart}, Adam~T. {Deller}, Keith~W. {Bannister}, Clancy~W. {James}, Chris {Flynn}, Cherie~K. {Day}, Shivani {Bhandari}, and Tara {Murphy}.
\newblock {A population analysis of pulse broadening in ASKAP fast radio bursts}.
\newblock {\em \mnras}, 497(2):1382--1390, September 2020.
\newblock \href {http://arxiv.org/abs/2006.16502} {\path{arXiv:2006.16502}}, \href {http://dx.doi.org/10.1093/mnras/staa1916} {\path{[DOI]}}, {\small[\href{https://ui.adsabs.harvard.edu/abs/2020MNRAS.497.1382Q}{ADS}]}.

\bibitem{Ocker2023}
Stella~Koch {Ocker}, James~M. {Cordes}, Shami {Chatterjee}, Di~{Li}, Chen-Hui {Niu}, James~W. {McKee}, Casey~J. {Law}, and Reshma {Anna-Thomas}.
\newblock {Scattering variability detected from the circumsource medium of FRB 20190520B}.
\newblock {\em \mnras}, 519(1):821--830, February 2023.
\newblock \href {http://arxiv.org/abs/2210.01975} {\path{arXiv:2210.01975}}, \href {http://dx.doi.org/10.1093/mnras/stac3547} {\path{[DOI]}}, {\small[\href{https://ui.adsabs.harvard.edu/abs/2023MNRAS.519..821O}{ADS}]}.

\bibitem{Bannister2019}
K.~W. {Bannister et al}.
\newblock {A single fast radio burst localized to a massive galaxy at cosmological distance}.
\newblock {\em Science}, 365(6453):565--570, August 2019.
\newblock \href {http://arxiv.org/abs/1906.11476} {\path{arXiv:1906.11476}}, \href {http://dx.doi.org/10.1126/science.aaw5903} {\path{[DOI]}}, {\small[\href{https://ui.adsabs.harvard.edu/abs/2019Sci...365..565B}{ADS}]}.

\bibitem{Ryder2023}
S.~D. {Ryder}, K.~W. {Bannister}, S.~{Bhandari}, A.~T. {Deller}, R.~D. {Ekers}, M.~{Glowacki}, A.~C. {Gordon}, K.~{Gourdji}, C.~W. {James}, C.~D. {Kilpatrick}, W.~{Lu}, L.~{Marnoch}, V.~A. {Moss}, J.~X. {Prochaska}, H.~{Qiu}, E.~M. {Sadler}, S.~{Simha}, M.~W. {Sammons}, D.~R. {Scott}, N.~{Tejos}, and R.~M. {Shannon}.
\newblock {A luminous fast radio burst that probes the Universe at redshift 1}.
\newblock {\em Science}, 382(6668):294--299, October 2023.
\newblock \href {http://arxiv.org/abs/2210.04680} {\path{arXiv:2210.04680}}, \href {http://dx.doi.org/10.1126/science.adf2678} {\path{[DOI]}}, {\small[\href{https://ui.adsabs.harvard.edu/abs/2023Sci...382..294R}{ADS}]}.

\bibitem{BK2025a}
Paz {Beniamini} and Pawan {Kumar}.
\newblock {Can repeating and non-repeating FRBs be drawn from the same population?}
\newblock {\em arXiv e-prints}, page arXiv:2506.09138, June 2025.
\newblock \href {http://arxiv.org/abs/2506.09138} {\path{arXiv:2506.09138}}, \href {http://dx.doi.org/10.48550/arXiv.2506.09138} {\path{[DOI]}}, {\small[\href{https://ui.adsabs.harvard.edu/abs/2025arXiv250609138B}{ADS}]}.

\bibitem{Gupta2025}
Om~{Gupta}, Paz {Beniamini}, Pawan {Kumar}, and Steven~L. {Finkelstein}.
\newblock {The cosmic evolution of FRBs inferred from CHIME/FRB Catalog 1}.
\newblock {\em arXiv e-prints}, page arXiv:2501.09810, January 2025.
\newblock \href {http://arxiv.org/abs/2501.09810} {\path{arXiv:2501.09810}}, \href {http://dx.doi.org/10.48550/arXiv.2501.09810} {\path{[DOI]}}, {\small[\href{https://ui.adsabs.harvard.edu/abs/2025arXiv250109810G}{ADS}]}.

\bibitem{CHIME2021}
Masoud {Rafiei-Ravandi}, Kendrick~M. {Smith}, Dongzi {Li}, Kiyoshi~W. {Masui}, Alexander {Josephy}, Matt {Dobbs}, Dustin {Lang}, Mohit {Bhardwaj}, Chitrang {Patel}, Kevin {Bandura}, Sabrina {Berger}, P.~J. {Boyle}, Charanjot {Brar}, Daniela {Breitman}, Tomas {Cassanelli}, Pragya {Chawla}, Fengqiu {Adam Dong}, Emmanuel {Fonseca}, B.~M. {Gaensler}, Utkarsh {Giri}, Deborah~C. {Good}, Mark {Halpern}, Jane {Kaczmarek}, Victoria~M. {Kaspi}, Calvin {Leung}, Hsiu-Hsien {Lin}, Juan {Mena-Parra}, B.~W. {Meyers}, D.~{Michilli}, Moritz {M{\"u}nchmeyer}, Cherry {Ng}, Emily {Petroff}, Ziggy {Pleunis}, Mubdi {Rahman}, Pranav {Sanghavi}, Paul {Scholz}, Kaitlyn {Shin}, Ingrid~H. {Stairs}, Shriharsh~P. {Tendulkar}, Keith {Vanderlinde}, and Andrew {Zwaniga}.
\newblock {CHIME/FRB Catalog 1 Results: Statistical Cross-correlations with Large-scale Structure}.
\newblock {\em \apj}, 922(1):42, November 2021.
\newblock \href {http://arxiv.org/abs/2106.04354} {\path{arXiv:2106.04354}}, \href {http://dx.doi.org/10.3847/1538-4357/ac1dab} {\path{[DOI]}}, {\small[\href{https://ui.adsabs.harvard.edu/abs/2021ApJ...922...42R}{ADS}]}.

\bibitem{2025Natur.637...48N}
Kenzie {Nimmo}, Ziggy {Pleunis}, Paz {Beniamini}, Pawan {Kumar}, Adam~E. {Lanman}, D.~Z. {Li}, Robert {Main}, Mawson~W. {Sammons}, Shion {Andrew}, Mohit {Bhardwaj}, Shami {Chatterjee}, Alice~P. {Curtin}, Emmanuel {Fonseca}, B.~M. {Gaensler}, Ronniy~C. {Joseph}, Zarif {Kader}, Victoria~M. {Kaspi}, Mattias {Lazda}, Calvin {Leung}, Kiyoshi~W. {Masui}, Ryan {Mckinven}, Daniele {Michilli}, Ayush {Pandhi}, Aaron~B. {Pearlman}, Masoud {Rafiei-Ravandi}, Ketan~R. {Sand}, Kaitlyn {Shin}, Kendrick {Smith}, and Ingrid~H. {Stairs}.
\newblock {Magnetospheric origin of a fast radio burst constrained using scintillation}.
\newblock {\em \nat}, 637(8044):48--51, January 2025.
\newblock \href {http://arxiv.org/abs/2406.11053} {\path{arXiv:2406.11053}}, \href {http://dx.doi.org/10.1038/s41586-024-08297-w} {\path{[DOI]}}, {\small[\href{https://ui.adsabs.harvard.edu/abs/2025Natur.637...48N}{ADS}]}.

\bibitem{2025arXiv250301740C}
Xiang-Lei {Chen}, Chao-Wei {Tsai}, Daniel {Stern}, Christopher~D. {Bochenek}, Shami {Chatterjee}, Casey {Law}, Di~{Li}, Chenhui {Niu}, Yuu {Niino}, Yi~{Feng}, Pei {Wang}, Roberto~J. {Assef}, Guodong {Li}, Sean~E. {Lake}, Gan {Luo}, and Mai {Liao}.
\newblock {The Host Galaxy of FRB 20190520B and Its Unique Ionized Gas Distribution}.
\newblock {\em arXiv e-prints}, page arXiv:2503.01740, March 2025.
\newblock \href {http://arxiv.org/abs/2503.01740} {\path{arXiv:2503.01740}}, \href {http://dx.doi.org/10.48550/arXiv.2503.01740} {\path{[DOI]}}, {\small[\href{https://ui.adsabs.harvard.edu/abs/2025arXiv250301740C}{ADS}]}.

\bibitem{Mahony2018}
Elizabeth~K. {Mahony}, Ron~D. {Ekers}, Jean-Pierre {Macquart}, Elaine~M. {Sadler}, Keith~W. {Bannister}, Shivani {Bhandari}, Chris {Flynn}, B{\"a}rbel~S. {Koribalski}, J.~Xavier {Prochaska}, Stuart~D. {Ryder}, Ryan~M. {Shannon}, Nicolas {Tejos}, Matthew~T. {Whiting}, and O.~I. {Wong}.
\newblock {A Search for the Host Galaxy of FRB 171020}.
\newblock {\em \apjl}, 867(1):L10, November 2018.
\newblock \href {http://arxiv.org/abs/1810.04354} {\path{arXiv:1810.04354}}, \href {http://dx.doi.org/10.3847/2041-8213/aae7cb} {\path{[DOI]}}, {\small[\href{https://ui.adsabs.harvard.edu/abs/2018ApJ...867L..10M}{ADS}]}.

\bibitem{Law2024}
Casey~J. {Law}, Kritti {Sharma}, Vikram {Ravi}, Ge~{Chen}, Morgan {Catha}, Liam {Connor}, Jakob~T. {Faber}, Gregg {Hallinan}, Charlie {Harnach}, Greg {Hellbourg}, Rick {Hobbs}, David {Hodge}, Mark {Hodges}, James~W. {Lamb}, Paul {Rasmussen}, Myles~B. {Sherman}, Jun {Shi}, Dana {Simard}, Reynier {Squillace}, Sander {Weinreb}, David~P. {Woody}, and Nitika~Yadlapalli {Yurk}.
\newblock {Deep Synoptic Array Science: First FRB and Host Galaxy Catalog}.
\newblock {\em \apj}, 967(1):29, May 2024.
\newblock \href {http://arxiv.org/abs/2307.03344} {\path{arXiv:2307.03344}}, \href {http://dx.doi.org/10.3847/1538-4357/ad3736} {\path{[DOI]}}, {\small[\href{https://ui.adsabs.harvard.edu/abs/2024ApJ...967...29L}{ADS}]}.

\bibitem{Amiri2025}
Mandana {Amiri et al}.
\newblock {A Catalog of Local Universe Fast Radio Bursts from CHIME/FRB and the KKO Outrigger}.
\newblock {\em arXiv e-prints}, page arXiv:2502.11217, February 2025.
\newblock \href {http://arxiv.org/abs/2502.11217} {\path{arXiv:2502.11217}}, {\small[\href{https://ui.adsabs.harvard.edu/abs/2025arXiv250211217A}{ADS}]}.

\bibitem{Shannon2024}
R.~M. {Shannon et al}.
\newblock {The Commensal Real-time ASKAP Fast Transient incoherent-sum survey}.
\newblock {\em arXiv e-prints}, page arXiv:2408.02083, August 2024.
\newblock \href {http://arxiv.org/abs/2408.02083} {\path{arXiv:2408.02083}}, \href {http://dx.doi.org/10.48550/arXiv.2408.02083} {\path{[DOI]}}, {\small[\href{https://ui.adsabs.harvard.edu/abs/2024arXiv240802083S}{ADS}]}.

\bibitem{Bhardwaj2024}
Mohit {Bhardwaj et al}.
\newblock {Host Galaxies for Four Nearby CHIME/FRB Sources and the Local Universe FRB Host Galaxy Population}.
\newblock {\em \apjl}, 971(2):L51, August 2024.
\newblock \href {http://arxiv.org/abs/2310.10018} {\path{arXiv:2310.10018}}, \href {http://dx.doi.org/10.3847/2041-8213/ad64d1} {\path{[DOI]}}, {\small[\href{https://ui.adsabs.harvard.edu/abs/2024ApJ...971L..51B}{ADS}]}.

\bibitem{Andersen2023}
{Chime/Frb Collaboration} and {Andersen et al}.
\newblock {CHIME/FRB Discovery of 25 Repeating Fast Radio Burst Sources}.
\newblock {\em \apj}, 947(2):83, April 2023.
\newblock \href {http://arxiv.org/abs/2301.08762} {\path{arXiv:2301.08762}}, \href {http://dx.doi.org/10.3847/1538-4357/acc6c1} {\path{[DOI]}}, {\small[\href{https://ui.adsabs.harvard.edu/abs/2023ApJ...947...83C}{ADS}]}.

\bibitem{Michilli2023}
Daniele {Michilli et al}.
\newblock {Subarcminute Localization of 13 Repeating Fast Radio Bursts Detected by CHIME/FRB}.
\newblock {\em \apj}, 950(2):134, June 2023.
\newblock \href {http://arxiv.org/abs/2212.11941} {\path{arXiv:2212.11941}}, \href {http://dx.doi.org/10.3847/1538-4357/accf89} {\path{[DOI]}}, {\small[\href{https://ui.adsabs.harvard.edu/abs/2023ApJ...950..134M}{ADS}]}.

\bibitem{Ravi2023}
Vikram {Ravi et al}.
\newblock {Deep Synoptic Array Science: Discovery of the Host Galaxy of FRB 20220912A}.
\newblock {\em \apjl}, 949(1):L3, May 2023.
\newblock \href {http://arxiv.org/abs/2211.09049} {\path{arXiv:2211.09049}}, \href {http://dx.doi.org/10.3847/2041-8213/acc4b6} {\path{[DOI]}}, {\small[\href{https://ui.adsabs.harvard.edu/abs/2023ApJ...949L...3R}{ADS}]}.

\bibitem{Ravi2022}
Vikram {Ravi}, Casey~J. {Law}, Dongzi {Li}, Kshitij {Aggarwal}, Mohit {Bhardwaj}, Sarah {Burke-Spolaor}, Liam {Connor}, T.~Joseph~W. {Lazio}, Dana {Simard}, Jean {Somalwar}, and Shriharsh~P. {Tendulkar}.
\newblock {The host galaxy and persistent radio counterpart of FRB 20201124A}.
\newblock {\em \mnras}, 513(1):982--990, June 2022.
\newblock \href {http://arxiv.org/abs/2106.09710} {\path{arXiv:2106.09710}}, \href {http://dx.doi.org/10.1093/mnras/stac465} {\path{[DOI]}}, {\small[\href{https://ui.adsabs.harvard.edu/abs/2022MNRAS.513..982R}{ADS}]}.

\bibitem{Ibik2024}
Adaeze~L. {Ibik et al}.
\newblock {Proposed Host Galaxies of Repeating Fast Radio Burst Sources Detected by CHIME/FRB}.
\newblock {\em \apj}, 961(1):99, January 2024.
\newblock \href {http://arxiv.org/abs/2304.02638} {\path{arXiv:2304.02638}}, \href {http://dx.doi.org/10.3847/1538-4357/ad0893} {\path{[DOI]}}, {\small[\href{https://ui.adsabs.harvard.edu/abs/2024ApJ...961...99I}{ADS}]}.

\bibitem{Heintz2020}
Kasper~E. {Heintz et al}.
\newblock {Host Galaxy Properties and Offset Distributions of Fast Radio Bursts: Implications for Their Progenitors}.
\newblock {\em \apj}, 903(2):152, November 2020.
\newblock \href {http://arxiv.org/abs/2009.10747} {\path{arXiv:2009.10747}}, \href {http://dx.doi.org/10.3847/1538-4357/abb6fb} {\path{[DOI]}}, {\small[\href{https://ui.adsabs.harvard.edu/abs/2020ApJ...903..152H}{ADS}]}.

\bibitem{Bhandari2022}
Shivani {Bhandari et al}.
\newblock {Characterizing the Fast Radio Burst Host Galaxy Population and its Connection to Transients in the Local and Extragalactic Universe}.
\newblock {\em \aj}, 163(2):69, February 2022.
\newblock \href {http://arxiv.org/abs/2108.01282} {\path{arXiv:2108.01282}}, \href {http://dx.doi.org/10.3847/1538-3881/ac3aec} {\path{[DOI]}}, {\small[\href{https://ui.adsabs.harvard.edu/abs/2022AJ....163...69B}{ADS}]}.

\bibitem{Gordon2023_1}
Alexa~C. {Gordon et al}.
\newblock {The Demographics, Stellar Populations, and Star Formation Histories of Fast Radio Burst Host Galaxies: Implications for the Progenitors}.
\newblock {\em \apj}, 954(1):80, September 2023.
\newblock \href {http://arxiv.org/abs/2302.05465} {\path{arXiv:2302.05465}}, \href {http://dx.doi.org/10.3847/1538-4357/ace5aa} {\path{[DOI]}}, {\small[\href{https://ui.adsabs.harvard.edu/abs/2023ApJ...954...80G}{ADS}]}.

\bibitem{Prochaska2019}
J.~Xavier {Prochaska}, Jean-Pierre {Macquart}, Matthew {McQuinn}, Sunil {Simha}, Ryan~M. {Shannon}, Cherie~K. {Day}, Lachlan {Marnoch}, Stuart {Ryder}, Adam {Deller}, Keith~W. {Bannister}, Shivani {Bhandari}, Rongmon {Bordoloi}, John {Bunton}, Hyerin {Cho}, Chris {Flynn}, Elizabeth~K. {Mahony}, Chris {Phillips}, Hao {Qiu}, and Nicolas {Tejos}.
\newblock {The low density and magnetization of a massive galaxy halo exposed by a fast radio burst}.
\newblock {\em Science}, 366(6462):231--234, October 2019.
\newblock \href {http://arxiv.org/abs/1909.11681} {\path{arXiv:1909.11681}}, \href {http://dx.doi.org/10.1126/science.aay0073} {\path{[DOI]}}, {\small[\href{https://ui.adsabs.harvard.edu/abs/2019Sci...366..231P}{ADS}]}.

\bibitem{Law2020}
Casey~J. {Law}, Bryan~J. {Butler}, J.~Xavier {Prochaska}, Barak {Zackay}, Sarah {Burke-Spolaor}, Alexandra {Mannings}, Nicolas {Tejos}, Alexander {Josephy}, Bridget {Andersen}, Pragya {Chawla}, Kasper~E. {Heintz}, Kshitij {Aggarwal}, Geoffrey~C. {Bower}, Paul~B. {Demorest}, Charles~D. {Kilpatrick}, T.~Joseph~W. {Lazio}, Justin {Linford}, Ryan {Mckinven}, Shriharsh {Tendulkar}, and Sunil {Simha}.
\newblock {A Distant Fast Radio Burst Associated with Its Host Galaxy by the Very Large Array}.
\newblock {\em \apj}, 899(2):161, August 2020.
\newblock \href {http://arxiv.org/abs/2007.02155} {\path{arXiv:2007.02155}}, \href {http://dx.doi.org/10.3847/1538-4357/aba4ac} {\path{[DOI]}}, {\small[\href{https://ui.adsabs.harvard.edu/abs/2020ApJ...899..161L}{ADS}]}.

\bibitem{Ravi2019}
V.~{Ravi}, M.~{Catha}, L.~{D'Addario}, S.~G. {Djorgovski}, G.~{Hallinan}, R.~{Hobbs}, J.~{Kocz}, S.~R. {Kulkarni}, J.~{Shi}, H.~K. {Vedantham}, S.~{Weinreb}, and D.~P. {Woody}.
\newblock {A fast radio burst localized to a massive galaxy}.
\newblock {\em \nat}, 572(7769):352--354, August 2019.
\newblock \href {http://arxiv.org/abs/1907.01542} {\path{arXiv:1907.01542}}, \href {http://dx.doi.org/10.1038/s41586-019-1389-7} {\path{[DOI]}}, {\small[\href{https://ui.adsabs.harvard.edu/abs/2019Natur.572..352R}{ADS}]}.

\bibitem{Gordon2023}
Alexa~C. {Gordon}, Wen-fai {Fong}, Sunil {Simha}, Yuxin {Dong}, Charles~D. {Kilpatrick}, Adam~T. {Deller}, Stuart~D. {Ryder}, Tarraneh {Eftekhari}, Marcin {Glowacki}, Lachlan {Marnoch}, August~R. {Muller}, Anya~E. {Nugent}, Antonella {Palmese}, J.~Xavier {Prochaska}, Marc {Rafelski}, Ryan~M. {Shannon}, and Nicolas {Tejos}.
\newblock {A Fast Radio Burst in a Compact Galaxy Group at $z$\raisebox{-0.5ex}\textasciitilde1}.
\newblock {\em arXiv e-prints}, page arXiv:2311.10815, November 2023.
\newblock \href {http://arxiv.org/abs/2311.10815} {\path{arXiv:2311.10815}}, \href {http://dx.doi.org/10.48550/arXiv.2311.10815} {\path{[DOI]}}, {\small[\href{https://ui.adsabs.harvard.edu/abs/2023arXiv231110815G}{ADS}]}.

\bibitem{Caleb2022}
Manisha {Caleb}, Ian {Heywood}, Kaustubh {Rajwade}, Mateusz {Malenta}, Benjamin~Willem {Stappers}, Ewan {Barr}, Weiwei {Chen}, Vincent {Morello}, Sotiris {Sanidas}, Jakob {van den Eijnden}, Michael {Kramer}, David {Buckley}, Jaco {Brink}, Sara~Elisa {Motta}, Patrick {Woudt}, Patrick {Weltevrede}, Fabian {Jankowski}, Mayuresh {Surnis}, Sarah {Buchner}, Mechiel~Christiaan {Bezuidenhout}, Laura~Nicole {Driessen}, and Rob {Fender}.
\newblock {Discovery of a radio-emitting neutron star with an ultra-long spin period of 76 s}.
\newblock {\em Nature Astronomy}, 6:828--836, May 2022.
\newblock \href {http://arxiv.org/abs/2206.01346} {\path{arXiv:2206.01346}}, \href {http://dx.doi.org/10.1038/s41550-022-01688-x} {\path{[DOI]}}, {\small[\href{https://ui.adsabs.harvard.edu/abs/2022NatAs...6..828C}{ADS}]}.

\bibitem{Beniamini2023}
P.~{Beniamini}, Z.~{Wadiasingh}, J.~{Hare}, K.~M. {Rajwade}, G.~{Younes}, and A.~J. {van der Horst}.
\newblock {Evidence for an abundant old population of Galactic ultra-long period magnetars and implications for fast radio bursts}.
\newblock {\em \mnras}, 520(2):1872--1894, April 2023.
\newblock \href {http://arxiv.org/abs/2210.09323} {\path{arXiv:2210.09323}}, \href {http://dx.doi.org/10.1093/mnras/stad208} {\path{[DOI]}}, {\small[\href{https://ui.adsabs.harvard.edu/abs/2023MNRAS.520.1872B}{ADS}]}.

\bibitem{CHIME2020}
B.~C. {CHIME/FRB Collaboration}, Andersen, K.~M. {Bandura}, M.~{Bhardwaj}, A.~{Bij}, M.~M. {Boyce}, P.~J. {Boyle}, C.~{Brar}, T.~{Cassanelli}, P.~{Chawla}, T.~{Chen}, J.~F. {Cliche}, A.~{Cook}, D.~{Cubranic}, A.~P. {Curtin}, N.~T. {Denman}, M.~{Dobbs}, F.~Q. {Dong}, M.~{Fandino}, E.~{Fonseca}, B.~M. {Gaensler}, U.~{Giri}, D.~C. {Good}, M.~{Halpern}, A.~S. {Hill}, G.~F. {Hinshaw}, C.~{H{\"o}fer}, A.~{Josephy}, J.~W. {Kania}, V.~M. {Kaspi}, T.~L. {Landecker}, C.~{Leung}, D.~Z. {Li}, H.~H. {Lin}, K.~W. {Masui}, R.~{McKinven}, J.~{Mena-Parra}, M.~{Merryfield}, B.~W. {Meyers}, D.~{Michilli}, N.~{Milutinovic}, A.~{Mirhosseini}, M.~{M{\"u}nchmeyer}, A.~{Naidu}, L.~B. {Newburgh}, C.~{Ng}, C.~{Patel}, U.~L. {Pen}, T.~{Pinsonneault-Marotte}, Z.~{Pleunis}, B.~M. {Quine}, M.~{Rafiei-Ravandi}, M.~{Rahman}, S.~M. {Ransom}, A.~{Renard}, P.~{Sanghavi}, P.~{Scholz}, J.~R. {Shaw}, K.~{Shin}, S.~R. {Siegel}, S.~{Singh}, R.~J. {Smegal}, K.~M. {Smith}, I.~H. {Stairs}, C.~M. {Tan}, S.~P. {Tendulkar}, I.~{Tretyakov},
  K.~{Vanderlinde}, H.~{Wang}, D.~{Wulf}, and A.~V. {Zwaniga}.
\newblock {A bright millisecond-duration radio burst from a Galactic magnetar}.
\newblock {\em \nat}, 587(7832):54--58, November 2020.
\newblock \href {http://arxiv.org/abs/2005.10324} {\path{arXiv:2005.10324}}, \href {http://dx.doi.org/10.1038/s41586-020-2863-y} {\path{[DOI]}}, {\small[\href{https://ui.adsabs.harvard.edu/abs/2020Natur.587...54T}{ADS}]}.

\bibitem{STARE2020}
C.~D. {Bochenek}, V.~{Ravi}, K.~V. {Belov}, G.~{Hallinan}, J.~{Kocz}, S.~R. {Kulkarni}, and D.~L. {McKenna}.
\newblock {A fast radio burst associated with a Galactic magnetar}.
\newblock {\em \nat}, 587(7832):59--62, November 2020.
\newblock \href {http://arxiv.org/abs/2005.10828} {\path{arXiv:2005.10828}}, \href {http://dx.doi.org/10.1038/s41586-020-2872-x} {\path{[DOI]}}, {\small[\href{https://ui.adsabs.harvard.edu/abs/2020Natur.587...59B}{ADS}]}.

\bibitem{Scholz2020}
Paul {Scholz} and {Chime/Frb Collaboration}.
\newblock {A bright millisecond-timescale radio burst from the direction of the Galactic magnetar SGR 1935+2154}.
\newblock {\em The Astronomer's Telegram}, 13681:1, April 2020.
\newblock {\small[\href{https://ui.adsabs.harvard.edu/abs/2020ATel13681....1S}{ADS}]}.

\bibitem{Beniamini2019}
Paz {Beniamini}, Kenta {Hotokezaka}, Alexander {van der Horst}, and Chryssa {Kouveliotou}.
\newblock {Formation rates and evolution histories of magnetars}.
\newblock {\em \mnras}, 487(1):1426--1438, July 2019.
\newblock \href {http://arxiv.org/abs/1903.06718} {\path{arXiv:1903.06718}}, \href {http://dx.doi.org/10.1093/mnras/stz1391} {\path{[DOI]}}, {\small[\href{https://ui.adsabs.harvard.edu/abs/2019MNRAS.487.1426B}{ADS}]}.

\end{thebibliography}
}

\appendix

\section{List of FRBs}
\label{app:FRB}
\begin{figure}[H]
 \centering
 \includegraphics[width=0.8\columnwidth]{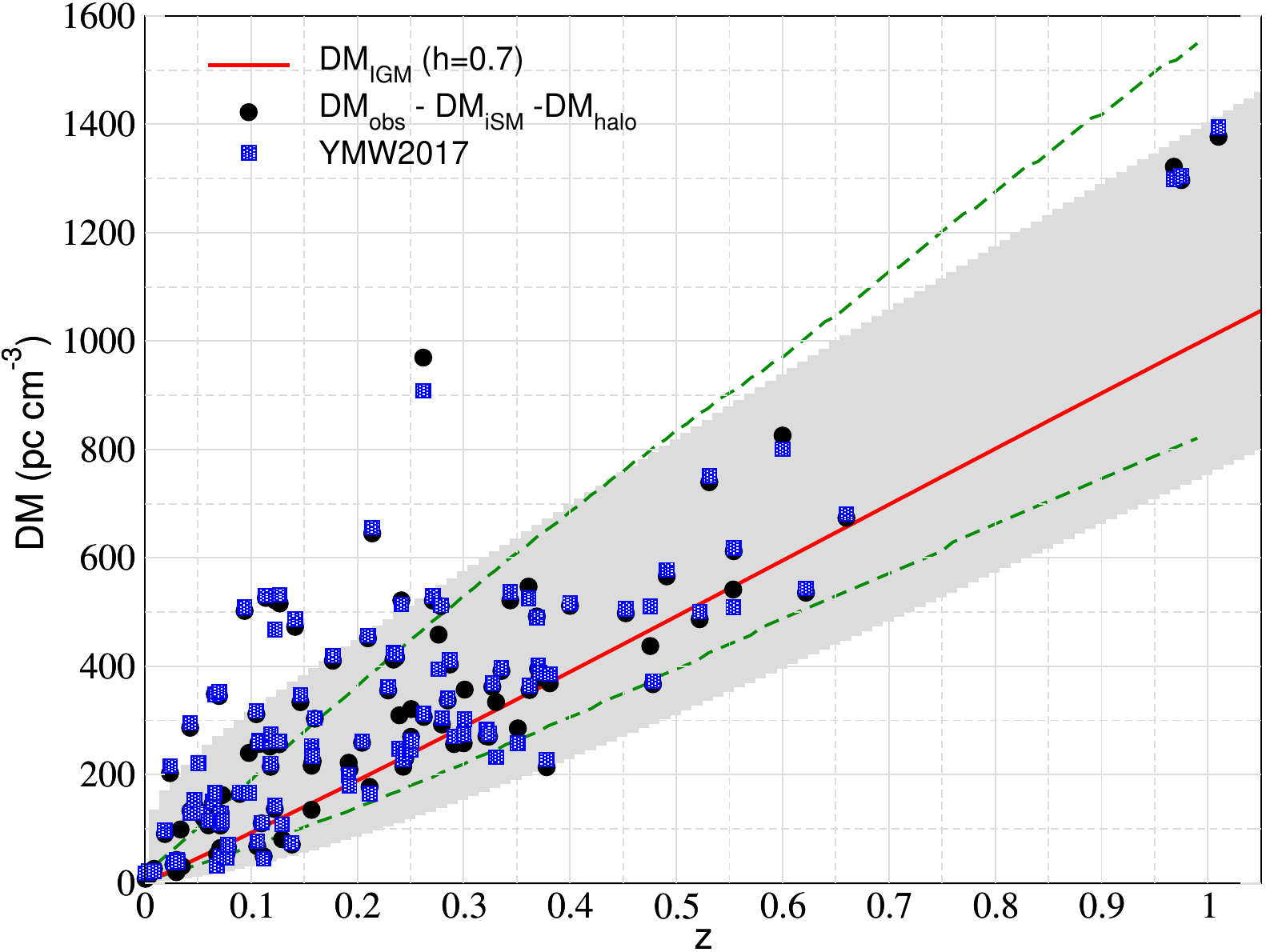}
 \caption{Dispersion measure of our sample of 110 FRBs after subtracting ISM and halo contribution form observed DM. We compare it to our fiducial IGM model with $h=0.7$. The  shaded grey region shows the 1$\sigma$ region due to scatter in IGM (Eq. \ref{eq:sigma_IGM}). We also show an estimate of the 68 percent confidence interval of ${\rm DM_{IGM}}$ distribution obtained from IllustrisTNG simulation \citep{Zhang2021} in green dashed lines.  }
 \label{fig:sample1}
\end{figure}

\begin{table}[H]
  \begin{center}
   \begin{tabular}{l|c|c|c|c|r} 
FRB & z & $\rm {DM_{obs}}$ & $\rm {DM_{ISM}}$ (NE2001) &  $\rm {DM_{ISM}}$ (YMW2017) & References \\
    \hline
    FRB20200120E & 0.0008 & 87.77 & 30 & 19.54 & \cite{Kirsten2022}  \\
    FRB20181030A & 0.0039 & 103.5 & 40 & 32 & \cite{Bhardwaj2021} \\
    FRB20171020A & 0.00867 & 114.1 & 38 & 26 & \cite{Mahony2018} \\    
   FRB20220319D & 0.0112 & 110.98 & 133.3 & 210.96 & \cite{Law2024}  \\
   FRB20231229A & 0.0190 & 198.5 & 58.12 & 51.78 & \cite{Amiri2025} \\
   FRB20240210A & 0.0238 & 283.75 & 31 & 17.90 & \cite{Shannon2024} \\
   FRB20181220A & 0.027 & 209.4 & 125.8 & 122.1 & \cite{Bhardwaj2024} \\
   FRB20231230A & 0.0298 & 131.4 & 61.51 & 83.24 & \cite{Amiri2025} \\
   FRB20181223C & 0.03024 & 112.51 & 19.91 & 19.1 & \cite{Bhardwaj2024} \\
   FRB20190425A & 0.03122 & 128.16 & 48.75 & 38.86 & \cite{Bhardwaj2024} \\
   FRB20180916B & 0.0337 & 348.76 & 200 & 324.95 & \cite{Marcote2020}  \\
   FRB20230718A & 0.035 & 477 & 396 & 467.25 & \cite{Shannon2024} \\
   FRB20240201A & 0.0427 & 374.5 & 38 & 29.15 & \cite{Shannon2024} \\
   FRB20220207C & 0.0430 & 262.38 & 79.3 & 83.27 & 
   \cite{Law2024}  \\
   FRB20211127I & 0.0469 & 234.83 & 42.5 & 31.46 &
   \cite{James2022}  \\
   FRB20201123A & 0.0507 & 433.55 & 251.93 & 162.4 & \cite{Rajwade2022} \\
   FRB20230926A & 0.0553 & 222.8 & 52.69 & 43.71 & \cite{Amiri2025} \\
   FRB20200223B & 0.06024 & 202.268 & 46 & 37 & \cite{Andersen2023} \\
   FRB20190303A & 0.064 & 222.4 & 26 & 21.79 & \cite{Michilli2023} \\
   FRB20231204A & 0.0644 & 221.0 & 29.73 & 21.79 & \cite{Amiri2025} \\
   FRB20231206A & 0.0659 & 457.7 & 59.13 & 59.29 & \cite{Amiri2025} \\
   FRB20210405I & 0.066 & 565.17 & 516.1 & 348.7 & \cite{Driessen2024} \\
   FRB20180814 & 0.068 & 189.4 & 87 & 108 & \cite{Michilli2023}
   \\
   FRB20231120A & 0.07 & 438.9 & 43.8 & 36.22 & \cite{Connor2024} \\
   FRB20231005A & 0.0713 & 189.4 & 33.37 & 28.79 & \cite{Amiri2025} \\
   FRB20190418A & 0.07132 & 184.5 & 70.1 & 85.6 & \cite{Bhardwaj2024} \\
   FRB20211212A & 0.0715 & 206.0 & 27.1 & 27.46 &
   \cite{James2022} \\
   FRB20231123A & 0.0729 & 302.1 & 89.76 & 136.89 & \cite{Amiri2025} \\
   FRB20220912A & 0.0771 & 219.46 & 115 & 122.24 & \cite{Ravi2023} \\
   FRB20231011A & 0.0783 & 186.3 & 70.36 & 65.70 & \cite{Amiri2025} \\   
   FRB20220509G & 0.0894 & 269.53 & 55.2 & 52.06 &
   \cite{Law2024} \\
   FRB20230124 & 0.0940 & 590.6 & 38.5 & 31.77 & \cite{Sharma2024},\cite{Connor2024} \\
   FRB20201124A & 0.098 & 413 & 123 & 196.67 & \cite{Ravi2022}  \\
   FRB20230708A & 0.105 & 411.51 & 50 & 43.90 & \cite{Shannon2024} \\
   FRB20231223C & 0.1059 & 165.8 & 47.9 & 38.64 & \cite{Amiri2025} \\
   FRB20191106C & 0.10775 & 333.4 & 25 & 21 & \cite{Ibik2024} \\
   FRB20231128A & 0.1079 & 331.6 & 25.05 & 20.54 & \cite{Amiri2025} \\
   FRB20230222B & 0.11 & 187.8 & 27.7 & 26.4 & \cite{Amiri2025} \\
   FRB20231201A & 0.1119 & 169.4 & 70.03 & 74.72 & \cite{Amiri2025} \\
   FRB20220914A& 0.1139 & 631.28 & 55.2 & 51.11 & \cite{Law2024} \\
   \end{tabular}
   \end{center}
   \end{table}
   \begin{table}[H]
  \begin{center}
   \begin{tabular}{l|c|c|c|c|r} 
      
   FRB & z & $\rm {DM_{obs}}$ & $\rm {DM_{ISM}}$ (NE2001) &  $\rm {DM_{ISM}}$ (YMW2017) & References \\
    \hline
     FRB20190608B & 0.1178 & 339 & 37 & 26.62 & \cite{Macquart2020} \\
   FRB20230703A & 0.1184 & 291.3 & 26.97 & 20.67 & \cite{Amiri2025} \\
   FRB20240213A & 0.1185 & 357.4 & 40.1 & 32.10 & \cite{Connor2024} \\
   FRB202030222A & 0.1223 & 706.1 & 134.13 & 188.14 & \cite{Amiri2025} \\
   FRB20190110C & 0.1224 & 221.961 & 35.66 & 28.96 & \cite{Ibik2024} \\
   FRB20230628A & 0.1265 & 345.15 & 39.1 & 33.36 & \cite{Sharma2024},\cite{Connor2024} \\
   FRB20240310A & 0.127 & 601.8 & 36 & 19.83 & \cite{Shannon2024} \\
   FRB20210807D & 0.1293 & 251.9 & 121.2 & 93.63 & \cite{James2022} \\
   FRB20240209A & 0.1384 & 176.49 & 55.5 & 52.2 & \cite{Shah2025} \\ 
   FRB20210410D & 0.1415 & 578.78 & 56.2 & 42.2 & \cite{Caleb2023} \\
   FRB20230203A &  0.1464 & 420.1 & 36.29 & 22.98 & \cite{Amiri2025} \\
   FRB20231226A & 0.1569 & 329.9 & 145 & 26.72 & \cite{Shannon2024} \\
   FRB20230526A & 0.157 & 316.4 & 50 & 21.88 & \cite{Shannon2024} \\
   FRB20220920A & 0.158 & 314.99 & 40.3 & 30.83 & \cite{Law2024} \\
   FRB20200430A & 0.1608 & 380.25 & 27 & 26.08 & \cite{Heintz2020} \\
      FRB20210603A & 0.177 & 500.15 & 40 & 30.79 & \cite{Cassanelli2024} \\
      FRB20230311A & 0.1918 & 364.3 & 92.39 & 115.68 & \cite{Amiri2025} \\
    FRB20220725A & 0.1926 & 290.4 & 31 & 61.24 & \cite{Shannon2024} \\
   FRB20221106A & 0.2044 & 343.8 & 35 & 31.85 & \cite{Shannon2024} \\
   FRB20240215A & 0.21 & 549.5 & 48.0 & 42.79 & \cite{Connor2024} \\
   FRB20230730A & 0.2115 & 312.5 & 85.18 & 97.29 & \cite{Amiri2025} \\
   FRB20210117A & 0.214 & 729.0 & 34.0 & 23.0 & \cite{Bhandari2023} \\
   FRB20221027A & 0.229 & 452.5 & 47.2 & 40.59 & \cite{Connor2024} \\
   FRB20191001A & 0.234 & 506.92 & 44.7 & 31.08 & \cite{Heintz2020}\\
   FRB20190714A & 0.2365 & 504.13 & 38 & 31.16 & \cite{Heintz2020} \\
   FRB20221101B & 0.2395 & 490.7 & 131.2 & 193.36 & \cite{Sharma2024},\cite{Connor2024} \\
   FRB20220825A & 0.2414 & 651.24 & 79.7 & 86.98 & \cite{Law2024} \\
   FRB20191228A & 0.2432 & 297.5 & 33 & 19.93 & \cite{Bhandari2022}\\
   FRB20231017A & 0.2450 & 344.2 & 64.55 & 55.64 & \cite{Amiri2025} \\ 
   FRB20221113A & 0.2505 & 411.4 & 91.7 & 115.40 & \cite{Sharma2024},\cite{Connor2024} \\
   FRB20220307B & 0.2507 & 499.15 & 128.2 & 186.98 & \cite{Sharma2024},\cite{Connor2024} \\
   FRB20220831A & 0.262 & 1146.25 & 126.7 & 188 & \cite{Connor2024} \\
   FRB20231123B & 0.2625 & 396.7 & 40.2 & 33.81 & \cite{Sharma2024},\cite{Connor2024} \\
   FRB20230307A & 0.2710 & 608.9 & 37.6 & 29.47 & \cite{Sharma2024},\cite{Connor2024} \\
   FRB20221116A & 0.2764 & 640.6 & 132.3 & 196.18 & \cite{Connor2024} \\
   FRB20220105A & 0.2785 & 583 & 22 & 20.64 & \cite{Shannon2024} \\
   FRB20210320C & 0.2796 & 384.8 & 42.2 & 30.39 & \cite{James2022} \\
    FRB20221012A & 0.2846 & 441.08 & 54.4 & 50.55 & \cite{Law2024} \\
   FRB20240229A & 0.287 & 491.15 & 37.9 & 29.52 & \cite{Connor2024} \\
   FRB20190102C & 0.2913 & 363.6 & 57.3 & 43.38 & \cite{Macquart2020} \\
   FRB20220506D & 0.3004 & 396.97 & 89.1 & 72.83 & \cite{Law2024} \\
     FRB20230501A & 0.3010 & 532.5 & 125.6 & 180.17 & \cite{Sharma2024},\cite{Connor2024} \\

   \end{tabular}
   \end{center}
   \end{table}

  \begin{table}[H]
  \begin{center}
   \begin{tabular}{l|c|c|c|c|r} 
      
   FRB & z & $\rm {DM_{obs}}$ & $\rm {DM_{ISM}}$ (NE2001) &  $\rm {DM_{ISM}}$ (YMW2017) & References \\
    \hline
       FRB20180924B & 0.3214 & 361.42 & 40.5 & 27.66 & \cite{Bannister2019} \\
   FRB20231025B & 0.3238 & 368.7 & 48.67 & 43.36 & \cite{Amiri2025} \\
   FRB20230626A & 0.3270 & 451.2 & 39.2 & 32.51 & \cite{Sharma2024},\cite{Connor2024} \\
     FRB20180301A & 0.3304 & 536 & 152 & 253.96 & \cite{Bhandari2022}\\
     FRB20231220A & 0.3355 & 491.2 & 49.9 & 44.54 & \cite{Connor2024} \\
   FRB20211203C & 0.3439 & 635.0 & 63.4 & 48.37 & \cite{Gordon2023_1} \\
   FRB20220208A & 0.3510 & 437.0 & 101.6 & 128.80 & \cite{Sharma2024},\cite{Connor2024} \\
   FRB20220726A & 0.3610 & 686.55 & 89.5 & 111.40 & \cite{Sharma2024},\cite{Connor2024} \\
   FRB20230902A & 0.3619 & 440.1 & 34 & 25.62 & \cite{Shannon2024} \\
   FRB20200906A & 0.3688 & 577.8 & 36 & 37.87 & \cite{Bhandari2022}\\
   FRB20240119A & 0.37 & 483.1 & 37.9 & 30.98 & \cite{Connor2024} \\
   FRB20220330D & 0.3714 & 468.1 & 38.6 & 30.09 & \cite{Sharma2024},\cite{Connor2024} \\
   FRB20190611B & 0.3778 & 321.4 & 57.8 & 43.67  &\cite{Heintz2020} \\
   FRB20220501C & 0.381 & 449.5 & 31 & 14 & \cite{Shannon2024} \\
   FRB20220204A & 0.4 & 612.2 & 50.7 & 46.03 & \cite{Connor2024} \\
   FRB20230712A & 0.4525 & 586.96 & 39.2 & 30.93 & \cite{Sharma2024}, \cite{Connor2024} \\
   FRB20181112A & 0.4755 & 589.27 & 42 & 29.03 & \cite{Prochaska2019} \\
   FRB20220310F & 0.4779 & 462.24 & 45.4 & 39.51 &\cite{Law2024} \\
   FRB20220918A & 0.491 & 656.8 & 41 & 28.85 & \cite{Shannon2024} \\
 FRB20190711A & 0.5220 & 593.1 & 56.4 & 42.62 &\cite{Heintz2020} \\
   FRB20230216A & 0.5310 & 828.0 & 38.5 & 27.05 & \cite{Sharma2024}, \cite{Connor2024} \\
   FRB20230814A & 0.5535 & 696.4 & 104.9 & 134.83 & \cite{Connor2024} \\
   FRB20221219A & 0.5540 & 706.7 & 44.4 & 38.60 & \cite{Sharma2024}, \cite{Connor2024} \\
   FRB20190614D & 0.60 & 959.2 & 83.5 & 108.72 & \cite{Law2020} \\
   FRB20220418A & 0.6220 & 623.25 & 37.6 & 29.54 & \cite{Law2024} \\
   FRB20190523A &  0.6600 & 760.8 & 37 & 29.88 & \cite{Ravi2019}\\
   FRB20240123A & 0.968 & 1462.0 & 90.3 & 112.98 & \cite{Connor2024} \\
   FRB20221029A & 0.9750 & 1391.05 & 43.9 & 36.4 & \cite{Sharma2024,Connor2024} \\
   FRB20220610A & 1.016 & 1458.1 & 30.9 & 13.58 & \cite{Gordon2023}

  \end{tabular}
   \caption{List of 110 FRBs used in this work along side their redshifts, total observed DM, ISM contribution using NE2001 and YMW2017 model and their published references. } 
   \label{tab:FRB_sample}
   \end{center}
   \end{table}


\section{Galactic radio magnetars}
\label{sec:Galmagnetars}
\color{black}We find 6 objects with a surface  magnetic field (inferred from $P,\dot{P}$ assuming dipole radiation) of $\gtrsim 10^{14}$ G, five of which are known magnetars \footnote{https://www.physics.mcgill.ca/~pulsar/magnetar/main.html} and the sixth is the long period transient and magnetar candidate, PSR J0901–4046 \citep{Caleb2022,Beniamini2023}.  
We also add the magnetar SGR 1935+2154 (the source of the Galactic FRB 20200428, \cite{CHIME2020,STARE2020}) which has a reported DM of $\approx 332$ pc cm$^{-3}$ \cite{Scholz2020}. The Galactic magnetar DMs are even larger than those of the youngest pulsars (this is not surprising as active magnetar ages are typically $\sim 10^4$\,yrs \cite{Beniamini2019}). Naively, the DMs of magnetars (which are of the order of 100-1000 pc cm$^{-3}$) may seem inconsistent with our inferred host DM. But from Fig. \ref{fig:pulsar_galactic}, we find them to be located at lower $b$ which largely explains these higher DMs. Therefore, the current data-set is consistent with magnetar progenitors of FRBs. \color{black}  

\end{document}